\documentclass[final,12pt,authoryear]{elsarticle}

\usepackage{amsmath}
\usepackage{amssymb}
\usepackage{amsfonts}
\usepackage{setspace}
\usepackage{rotating}
\usepackage{multirow}
\usepackage{graphicx}
\usepackage{color}
\usepackage{xcolor}
\usepackage{lineno}
\usepackage{framed}
\usepackage[all]{xy}
\usepackage{rotating}
\usepackage{verbatim}
\usepackage{graphicx}
\usepackage{subcaption}

\newcommand{\Real}{\mathbb R}

\newcommand{\x}{{\mathbf x}}

\newcommand{\cut}[1]{} %

\newcommand{\blue}[1]{\textcolor[rgb]{0,0,0.8}{#1}}           %

\usepackage{hyperref}
\hypersetup{
    unicode=false,          %
    pdftoolbar=true,        %
    pdfmenubar=true,        %
    pdffitwindow=false,     %
    pdfstartview={FitH},    %
    pdftitle={Crops},         %
    pdfauthor={Mateo et al}, %
    pdfsubject={Crops},       %
    pdfcreator={Mateo et al},  %
    pdfproducer={Mateo et al}, %
    pdfkeywords={keyword1} {key2} {key3}, %
    pdfnewwindow=true,      %
    colorlinks=true,       %
    linkcolor={black},          %
    citecolor=gray,        %
    filecolor=gray,      %
    urlcolor=gray           %
}

\journal{Remote Sensing of Environment}

\begin{document}

\begin{frontmatter}
\title{Synergistic Integration of Optical and Microwave Satellite Data for Crop Yield Estimation
\footnote{Paper published at Remote Sensing of Environment Volume 234, 1 December 2019, 111460, doi: https://doi.org/10.1016/j.rse.2019.111460}}

\author[uv]{Anna Mateo-Sanchis\corref{cor1}}
\ead{anna.mateo@uv.es}
\author[uv]{Maria Piles}
\author[uv]{Jordi Mu{\~n}oz-Mar\'i}
\author[uv]{Jose E. Adsuara}
\author[uv]{Adri\'an P\'erez-Suay}
\author[uv]{Gustau Camps-Valls}

\cortext[cor1]{Corresponding author}
\address[uv]{Image Processing Laboratory (IPL), Parc Cient\'ific, Universitat de Val\`encia, \newline
C/ Catedr\'atico Jos\'e Beltr\'an, 2. 46980 Paterna, Val\`encia, Spain. \newline
\url{anna.mateo@uv.es}, \url{http://isp.uv.es}}

\begin{abstract}
Developing accurate models of crop stress, phenology and productivity is of paramount importance, given the increasing need of food. Earth observation (EO) remote sensing data provides a unique source of information to monitor crops in a temporally resolved and spatially explicit way. In this study, we propose the combination of multisensor (optical and microwave) remote sensing data for crop yield estimation and forecasting using two novel approaches. We first propose the lag between Enhanced Vegetation Index (EVI) derived from MODIS and Vegetation Optical Depth (VOD) derived from SMAP as a new joint metric combining the information from the two satellite sensors in a unique feature or descriptor. Our second approach avoids summarizing statistics and uses machine learning to combine full time series of EVI and VOD.
This study considers \blue{two statistical methods, a regularized linear regression and its nonlinear extension called kernel ridge regression} to directly estimate the county-level surveyed total production, as well as individual yields of the major crops grown in the region: corn, soybean and wheat. The study area includes the US Corn Belt, and we use agricultural survey data from the National Agricultural Statistics Service (USDA-NASS) for year 2015 for quantitative assessment.
Results show that (1) the proposed EVI-VOD lag metric correlates well with crop yield and outperforms common single-sensor metrics for crop yield estimation; (2) the statistical (machine learning) models working directly with the time series largely improve results compared to previously reported estimations; (3) the combined exploitation of information from the optical and microwave data leads to improved predictions over the use of single sensor approaches with coefficient of determination R$^2\geq 0.76$; (4) when models are used for within-season forecasting with limited time information, crop yield prediction is feasible up to four months before harvest (models reach a plateau in accuracy); and (5) the robustness of the approach is confirmed in a multi-year setting, reaching similar performances than when using single-year data.
In conclusion, results confirm the value of using both EVI and VOD at the same time, and the advantage of using automatic machine learning models for crop yield/production estimation.
\end{abstract}


\end{frontmatter}

\clearpage
\begin{framed}
\section*{Highlights}
\begin{enumerate}
\item Optical and passive microwave data are complementary and useful to estimate crop yield
\item The time lag between EVI and VOD is proposed as a new metric for crop assessment
\item The EVI/VOD lag outperforms common single-sensor metrics for crop yield estimation
\item Machine learning further improves results by directly blending multisensor time series.
\item Crop yield prediction is feasible up to four months before harvest.
\end{enumerate}
\end{framed}

\newpage

\section{Introduction}
The increasing world demand in agricultural production is putting more and more pressure on the agroecosystems and on the food chain in general, bringing a new scenario for agricultural policies and scientific research. Currently, 7.6 billion people live on Earth and is expected to reach 9.8 billion people by 2050 \citep{mateo_sagasta18}. %
This increase will require a 70-100\% rise in food production given projected trends in diets, consumption, and income \citep{wart13}, and a regular increase in the area allocated for cultivation \citep{fieuzal17}. It also poses a challenge on food security monitoring systems \citep{foley11,you17}. In this regard, operational crop yield forecasting systems have critical value with a direct economic and social impact, from food prices to food policies, international trades, and land use decisions \citep{macdonald80, bauer81, Prasad06,lobell07, lopez15}. Accurate and timely crop yield estimation is currently one of the major challenges in agricultural research and of paramount interest to governments, public administrations, and farm managers \citep{ idso77, marinkovic09, Fritz2019}.

Traditionally, crop yield estimates were based on agro-meteorological models or on the compilation of the survey information that farmers provided during the growing season. Constraining agro-meteorological models is complicated, and often reconciliation with field observations is not possible. In addition, they are not generally adapted to larger scales and are time and resource consuming \citep{kastens05, wart13}. In the last decade, the availability of
Earth observation (EO) data has opened new ways for efficient agricultural mapping, crop monitoring and assessment. EO satellites naturally capture key information about crops that can influence yields with accuracy and timeliness. In this regard, it is a unique means to provide information about crop status over large areas with regular revisits, and allows deriving spatially explicit and temporally resolved maps of production and yield %
\citep{atzberger13,fieuzal17}.
Most studies on the use EO data for crop estimation are centered on visible and infrared sensors such as AVHRR, Landsat or MODIS. The information obtained in the visible range is most often associated to vegetation greenness (or chlorophyll content) and indirectly to plant growth, while infrared sensors are mainly used to examine the thermal characteristics of the vegetation \citep{gonzalez14, Fritz2019}. However, the temporal availability of visible and infrared EO data is problematic since, at these frequencies, the clouds prevent measuring the land surface, and therefore crop properties at key stages may not be recorded. Alternatively, some recent works have introduced the use of low-frequency passive microwave data for crop assessment  \citep{ patton13,  Hornbuckle2016, mladenova17, piles17, chaparro18, lewis_beck_18}. Passive microwave EO data provides information of the water content in soils and vegetation, even under cloudy conditions and at night. Still, the spatial resolution of passive microwave EO data is coarse ($\approx 30$km), and therefore they are most appropriate for large-scale crop yield studies.

In this work, we are interested in the synergistic use of optical and passive microwave data for the particular problem of crop yield estimation. To do so, analysis and fusion of both types of information is needed:
\begin{itemize}
    \item {\em Optical vegetation indices.} %
The great majority of studies summarize the multispectral information into a single metric known as vegetation indices (VIs) related to plant status~\citep{johnson14, Xiaocheng17}.
Actually, optical vegetation indices are easy to compute and useful to monitor the quantity, quality and behavior of the vegetation \citep{zhang2003}. The vegetation dynamics or phenology characterizes the seasonal timing of vegetation growing seasons, canopy growth and senescence \citep{jones11}. Among the most widely used VIs, the Normalized Difference Vegetation Index (NDVI) has been extensively and successfully used in agricultural mapping and monitoring, as well as in many crop yield studies \citep{quarmby93,  mkhabela11, bolton13, Chen18, Kern18}. In the majority of cases, NDVI is complemented with the Enhanced Vegetation Index (EVI) as it generally improves the sensitivity over biomass regions, overcomes the saturation problem of NDVI, and alleviates the influence of atmospheric effects \citep{zhang14, son14,zhong16}. In either case, the rationale for using VIs or optically-derived biophysical parameters such as the Leaf Area Index (LAI) and the fraction of photosynthetically active radiation (fAPAR), is that they are sensitive to the amount of photosynthetic active vegetation, which depends on the biotic and abiotic conditions that affect crop status and, ultimately, determine final yield \citep{LopezLozano15}. Accordingly, regression models have been proposed to relate VIs computed at a particular the day of year (DOY) with the yield \citep{doraiswamy03,bolton13}. Other studies used a shape model fitting derived from time series of VIs to detect the phenological stages of crops and determine the dates of the required input data \citep{sakamoto13, sakamoto14}. Interestingly, the anomalies of VIs during the growing season have also been used to predict changes in crop yield \citep{zhang10}. But commonly, the dominant approach integrates a single or multi-year time series of VIs with statistical regression models either in the form of seasonal or annual metrics \citep{ zhang03, li07, ren08}. %

\item {\em Microwave vegetation optical-depth.}
The vegetation optical depth (VOD) is a parameter accounting for the attenuation of microwave signals naturally emitted by the soil as they pass through the vegetation canopy \citep{Ulaby2014}. It is sensitive to the canopy structure and the amount of living biomass, being directly proportional to the volumetric water content of vegetation %
\citep{ Jackson1991, grant16, alemu17}. Also, it is relatively insensitive to signal degradation from solar illumination and atmospheric effects, which makes them an ideal piece of information for phenology and crop production studies \citep{jones11}.
Unlike optical VIs, which are related to the amount of green leaf biomass and constitute good indicators of photosynthetic activity,
the VOD accounts for other plant hydraulic properties and for the canopy structure to different degrees, depending on the frequency. VOD from longer microwave wavelengths (i.e., L-band) are able to penetrate the vegetation cover up to high densities and represent the canopy water in total above‐ground biomass \citep{tian16,guan17}. An environmental factor as essential as water can drive yield production in crops, and is therefore indispensable to take into account throughout crop growth \citep{idso77, mladenova17}. However, the applicability of VOD in agriculture, yet providing new venues of research, has been assessed in only few previous studies so far, most probably due to the coarse spatial resolution of current estimates %
\citep{guan17}.
\end{itemize}
Blending these two types of information is not new. Previous studies investigated relationships between VOD and different optical vegetation indices \citep{Liu_2011,lawrence14,piles17} and demonstrated that VOD provides new information about crop phenology that standard vegetation indices do not capture, and that it is able to follow the crop progress an their phenological phases \citep{Hornbuckle2016, chaparro18, lewis_beck_18}. The information captured by the two vegetation products is different and complementary \citep{clevers96, tian16}, and their synergy contributes to a more comprehensive view of land surface phenology \citep{jones11}. %
The fact is that optical-infrared VIs and VOD do not necessarily respond identically to changes in crop status \citep{Liu_2011}. Some studies combined both spectrum ranges and showed that VOD seasonal variations on VIs and VOD are highly synchronous, with the peak of VOD succeeding the peak of the VIs, with distinct time lags depending on the specific vegetation species or land use \citep{lawrence14, alemu17, tian18}, thus indicating that they account for different yet coupled processes. Indeed, the water stored in crop tissue increases through the growing vegetation stages and decreases during senescence \citep{patton13, Hornbuckle2016, piles17}. This general behaviour of the VOD signal allows to relate the phenology of the vegetation and yield in a variety of agro-ecosystems. Studies also demonstrated that VOD remained noticeably higher than NDVI throughout the non-growing season over sites dominated by mixed croplands \citep{jones11}. In general, the use of satellite data across various spectral ranges has been shown to improve monitoring of large-scale crop growth and yield beyond what can be achieved from individual sensors \citep{guan17,fieuzal17}.

In this study we introduce two approaches to synergistically combine optical and microwave data for the crop yield estimation problem:
\begin{itemize}
\item {\em Synergistic metric approach.} We propose the lag of maximum correlation between EVI and VOD time series as a novel joint metric for crop estimation. The proposed metric combines both time series in a summarizing statistic of the two time series, unlike standard metrics that focus only on summarizing individual time series.
\item {\em Machine learning approach.  } Summarizing metrics are practical and widely-used, but they constitute a quite limited approach: parametric models impose a simple and arbitrary form to the relation between the indices and the crop yield, and they disregard nonlinear feature relations. Alternatively, we also propose machine learning (ML) regression models that exploits the whole time series as feature vectors. ML models fit a non-parametric nonlinear model which typically improves the results.
\end{itemize}
We introduce the use of both a regularized linear regression (RLR) model and its nonlinear extension, called kernel ridge regression (KRR)~\citep{CampsValls09wiley}, to relate EVI and VOD time series to county-scale crop yield. %
Models are trained on agricultural survey data from the National Agricultural Statistics Service (USDA-NASS). %
We develop four models: a general crop production model, and three crop specific models for corn, soybean and wheat. Models are evaluated in scenarios of crop yield {\em estimation}, where all observations before the harvesting time are used, and in a {\em within-season forecasting} setting where an increasing number of time observations are included.
In both settings, we show empirically that (1) the proposed synergistic lag metric improves results over common single-sensor metrics, (2) the combination of optical and microwave data leads to largely improved predictions in all cases and crops, and %
(3) the machine learning models excel in exploiting multisource and multitemporal information. %
We also perform a dedicated experiment with multi-year data to assess the capability of the developed models to work with additional data.

\section{Data collection} \label{sec:data}
\subsection{Study Area} \label{studyareasec}

\begin{figure}[t!]
\centering
	\includegraphics[scale= 0.5]{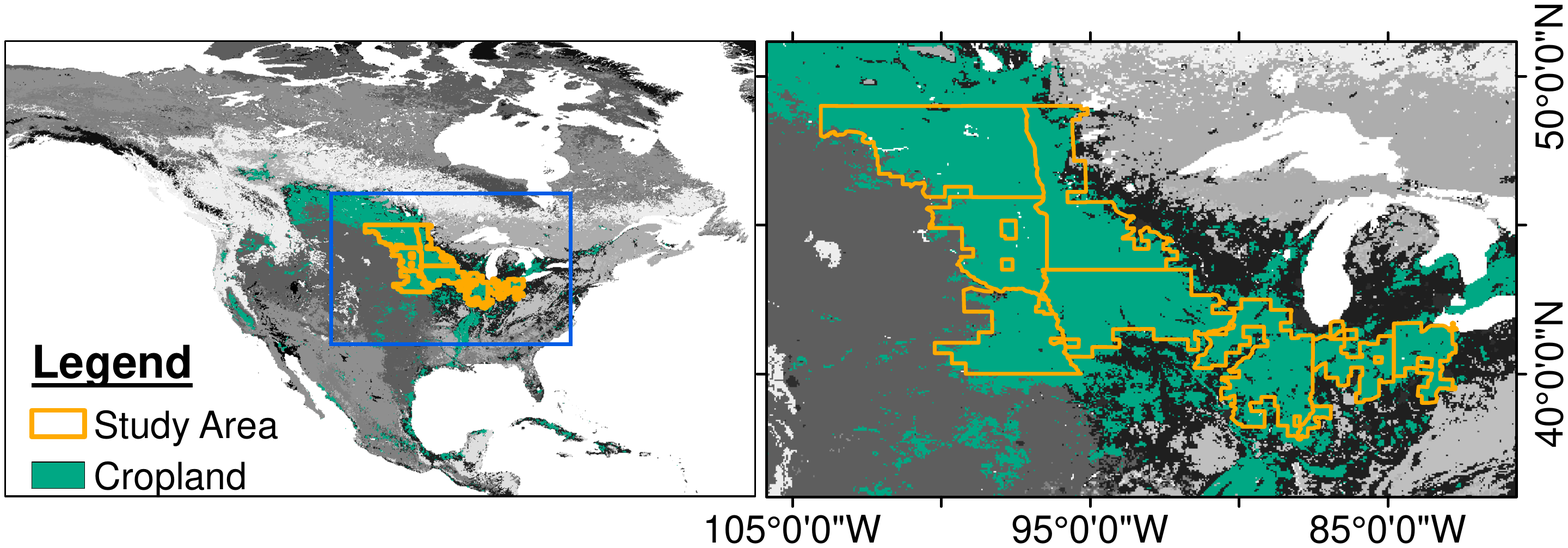} \\
	\vspace{-5.3cm}
    \caption{Study Area including the 8 states and cropland mask following the MODIS IGBP land cover classification. Only pixels classified as croplands were considered in this study. }
    \label{fig:studyarea9}
\end{figure}

The study is focused in the extensive croplands within the so-called Corn Belt of Midwestern United States (see Fig. \ref{fig:studyarea9}). It comprises eight states: Illinois, Indiana, Iowa, Minnesota, Nebraska, North Dakota, Ohio and South Dakota.
About 50\% of the annual production of corn and soybean of total U.S. production comes from the Corn Belt \citep{schnitkey13}, which makes it one of the world's most important food baskets \citep{zhong16, guan17}. For year 2015 (our study period), the United States had about  44\% of corn, 35\% of soybean and 10\% of wheat global production (FAO, \url{http://www.fao.org/home/en}).

Remarkably, the US Department of Agriculture (USDA) routinely collects and makes available extensive information on crop development and production over this region with a unique level of detail that allows the development of large-scale crop yield assessment algorithms based on satellite data. Hence, numerous studies and field experiments have been conducted in the U.S. Corn Belt for agricultural research and, more specifically, for crop estimation and prediction {e.g., \cite{zhang10,xin13,johnson14,guan17,chaparro18}.

\subsection{Survey data} \label{survey data}
The USDA's National Agricultural Statistics Service (USDA-NASS) conducts hundreds of surveys every year and prepares reports with detailed agricultural information covering the whole national territory (available at  \url{http://quickstats.nass.usda.gov/}). An exhaustive database can be queried to access information such as area planted, area harvested and yield per crop type planted at different geographical levels: country, state, agricultural district or county scales. Information on crop progress or \textit{phenology} is only available at the state level.

In this work, we selected county-level yield and area planted for the 2015 season for all the cultivated crops in the region of study, namely corn, soybeans, wheat (including winter, durum and spring varieties), oats, beans, barley, peas, canola, flaxseed, sorghum and lentils. Yield values per crop (given in bushels/acre or lb/acre, depending on the crop) were converted to kg$\cdot$m$^{-2}$ using published conversion tables \citep{usda1992weights}. The total area planted per county was obtained and used to compute the percentage of area planted per crop within each county, hereinafter referred to as `crop proportion.' A single yield value per county was then obtained as a weighted average of the yield and area planted of all the crops reported.

After screening out the counties with no EVI or VOD satellite data available for the study period, the agricultural survey dataset contained a total of 385 counties. Fig. \ref{fig:totalhist}a shows the histogram of the reported yield per county for year 2015 and for the three major crops grown in the region: corn, soybean and wheat. It shows that corn is the most productive crop (median yield of 1.13 kg$\cdot$m$^{-2}$), followed by wheat (median yield of 0.41 kg$\cdot$m$^{-2}$) and soybean (median yield of 0.37 kg$\cdot$m$^{-2}$). Regarding crop distribution and heterogeneity, percentages of area planted for the three crop types are displayed in Fig. \ref{fig:totalhist}b. Corn and soybean are the major crops in the central and southeastern areas. In contrast, wheat predominates across the north and northwestern areas, where other species are also present but in a lower proportion. Corn and soybean were planted in 363 and 361 counties, respectively, and wheat was planted in 204 counties.

\begin{figure}[b!] \centering
\centerline{}
	\includegraphics[height = 5cm]{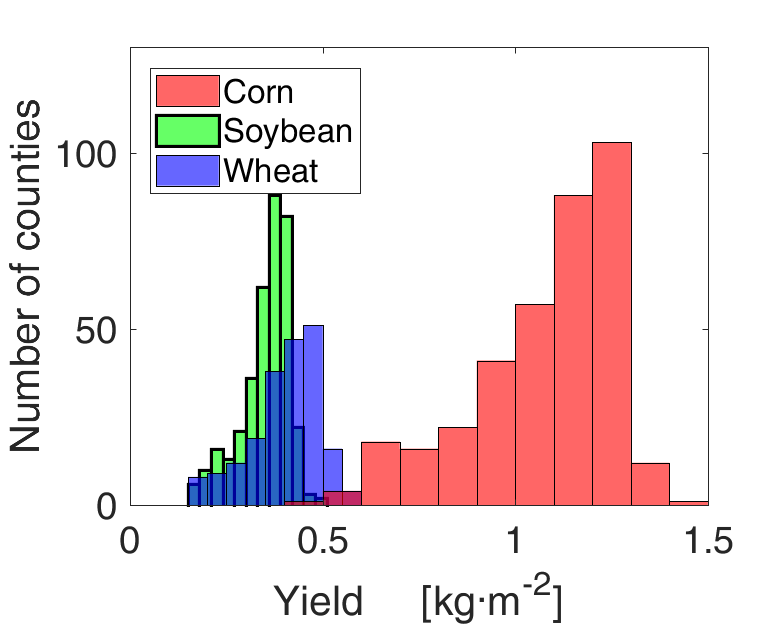}~
	\includegraphics[scale = 0.6]{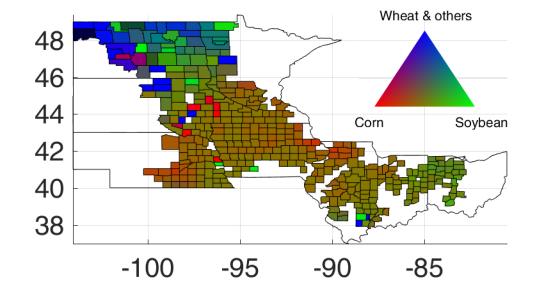}
    \caption{Left: histogram on the reported yield per county for the three major crops grown in the study area (soybean, wheat and corn). Right: County-scale percentages of area planted for corn, soybean, and other crops (mainly wheat). Each vertex in the triangle corresponds to 100\%}
    \label{fig:totalhist}
\end{figure}

\subsection{VOD data from SMAP} \label{SMAP VOD}

Microwave-based VOD is a measure of the attenuation of soil microwave emissions when they pass through the vegetation canopy \citep{Ulaby2014}. It is sensitive to the canopy structure and the amount of living biomass, being directly proportional to the volumetric water content of vegetation \citep{Jackson1991,Wigneron2017,Momen2017}. A great advantage with VOD is that, unlike optical vegetation indices, it is possible to recover information from microwaves through clouds. This is of considerable interest for agricultural applications in order not to miss important crop stages, because generally, almost two-thirds of the globe is covered with clouds. Here VOD is used for crop yield estimation alone or in combination with EVI. The VOD data set is obtained using the multitemporal dual channel algorithm (MT-DCA) \citep{Konings2016}. The algorithm was designed to estimate soil moisture and VOD from single look-angle L-band radiometric observations using two consecutive overpasses and no ancillary information on vegetation. SMAP VOD datasets retrieved using MT-DCA have shown good agreement with vegetation and land cover patterns at a variety of ecosystems \citep{piles17,Konings2017,chaparro18,Feldman2018}.

In this study, SMAP L-band VOD daily maps over the U.S Corn Belt (April to October 2015, 213 images) were used. The dataset is provided in the EASE 9 km grid \citep{Konings2017}. The following pre-processing steps were performed to the VOD time series: (i) pixels with frozen soils were screened out using the SMAP radiometer-based snow and frozen ground flags \citep{ONeill2018}. (ii) a 7-day moving average was applied to reduce high frequency noise and any temporal gaps in the time series were filled by fitting an auto-regressive model; the order of the model was selected iteratively on a pixel basis to minimize the Akaike information criterion ~\citep{Akaike69}. (iii) pixels with less than 19 days between minimum and maximum VOD were discarded, as suggested by \cite{chaparro18}.

\subsection{EVI data from MODIS}

The EVI is a VI/NIR index used as a proxy of vegetation condition and photosynthetic activity \citep{Huete2002}. Here it is used as input for crop yield estimation alone or in combination with microwave VOD. The EVI dataset is obtained from the MODIS/Terra MOD13C1 v6. product. It provides global 16-day composites on a 0.05$^\circ$ regular grid that were resampled in this work to the EASE2 9km grid using bilinear interpolation. A total of thirteen images were acquired for the period April to October 2015. MODIS products are available through NASA's Land Processes Distributed Active Archive Center (\href{https://lpdaac.usgs.gov/dataset_discovery/modis/modis_products_table}{LPDAAC}).

\subsection{Land cover} \label{land cover}
The MODIS International Geosphere-Biosphere Programme land cover (IGBP, MCD12C1 product v.6) was used here to identify the croplands in the region. We converted the classification map from its native 0.05$^\circ$ grid to the coarser EASE2 9 km grid using the statistical mode as a criteria for the spatial aggregation. Subsequently, we created a mask of purely cropland pixels and screened out satellite data from mixed and non-agricultural pixels from the study.

\section{Methodology} \label{sec:methods}
In this section we present the proposed methodology for crop yield estimation and forecasting using summarizing metrics and full time series of EVI and VOD. A general scheme of the study logic is shown in Fig.~\ref{fig:diagramSETUP}. In the next subsections we describe one-by-one the steps of the methodology.

\begin{figure}[t!]
    \centerline{\includegraphics[scale = 0.35]{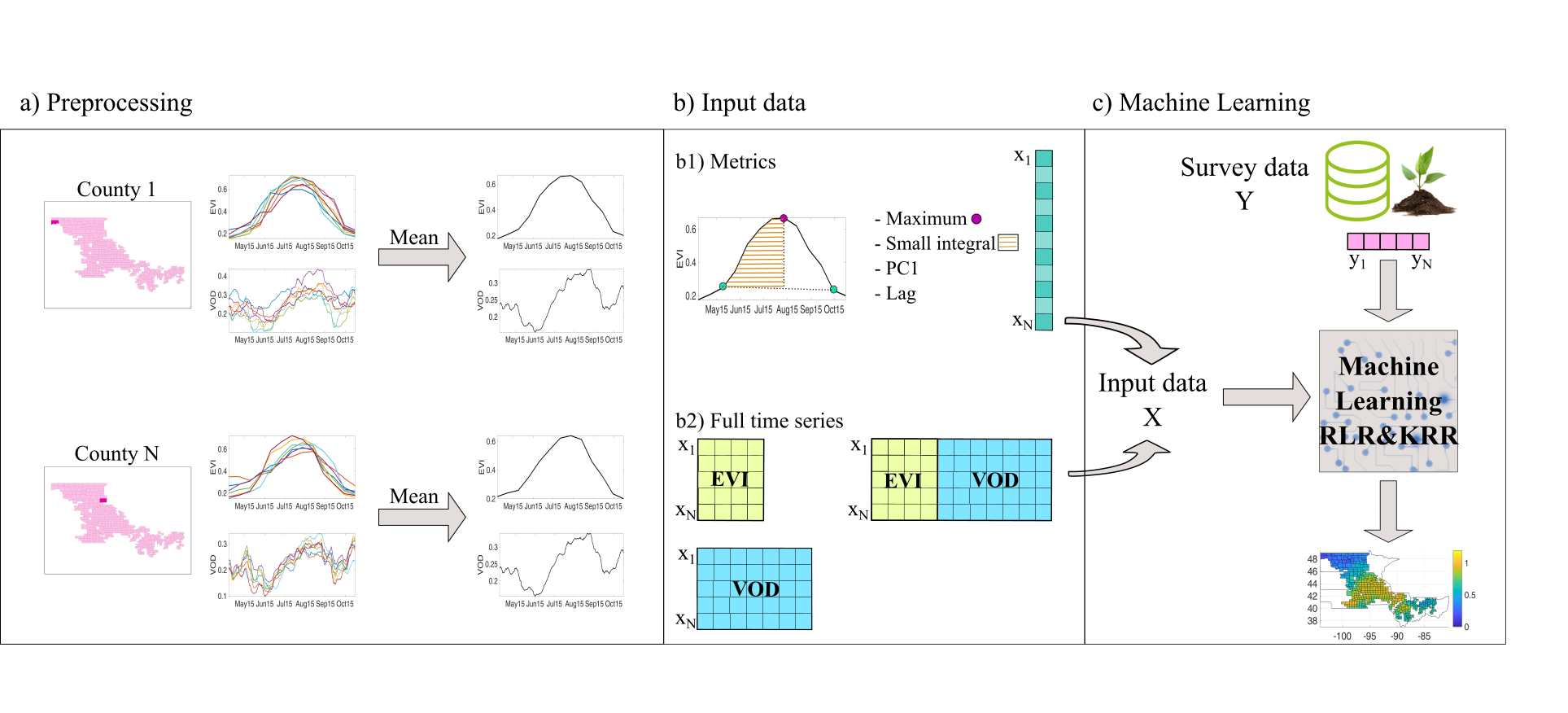}}
    \vspace{-1cm}
\caption{Schematic representation of the proposed methodology: a) pre-processing steps, b) set-up of input feature space, and c) crop yield assessment. We start by extracting the MODIS EVI and SMAP VOD time series at the county scale. These time series are then used to capture crop phenology and predict yield in two complementary ways: using summarizing metrics (i.e., maximum value, small integral, PC1 or EVI/VOD lag) or the full temporal series. Subsequently, these descriptors are fed into a machine learning regression algorithm and trained to estimate crop yield. After training the models with a reduced set, they are applied to the whole area to generate spatially explicit crop yield estimates.}
\label{fig:diagramSETUP}
\end{figure}

\subsection{Preprocessing}

To carry out the crop estimation experiments, we fixed a satellite temporal observational window from April to October 2015. This period includes the growing and senescence stages of all crops in the study area, and ensures estimations can be performed before harvest.
Usually, the temporal window is determined by the crop phenological stages, which depends on each crop type. This, however, requires the identification of the start/end of the photosynthetically active growing season to define the window, which is tedious, depends on the timing of greenup of the mixture of crops within each county, and in the end does not allow us to account for the variability in crop phenology. Instead, with the temporal window defined, we are able to use all data available within those seven months. %
Having the 7 months of EVI and VOD data projected in the 9 km EASE2 grid (see Section \ref{sec:data}), we need to relate the satellite pixel-based information to the survey data at the county scale. To do so, we extracted the pixels from each county according to its geographic boundaries given by shapefile polygons. Pixels that overlapped multiple counties were assigned only to the county enclosing the pixel centroid. Then, the pixels assigned to each county were spatially averaged to produce time series of EVI and VOD data at the county-level (Fig. \ref{fig:diagramSETUP} (a)). We used those county-averaged time series (and their derived metrics) directly as predictors. Note that an agricultural mask was applied to the satellite data, but no further classification per crop-type was performed. %

The result of the county-based aggregation was a $13$-dimensional and a $213$-dimensional feature vector per county corresponding to the EVI and VOD spatially-averaged time series, respectively. This allowed us to combine sensor information easily by just stacking the two feature vectors in a $226$-dimensional joined vector per county.
Alternatively, one can work with a summarizing metric instead of the full time series. In that case, only the metric (e.g., maximum and mean value, amplitude, integral, or our proposed lag metric) forms the feature vector summarizing the county information. A graphical representation of the two approaches is shown in Fig.~\ref{fig:diagramSETUP} (b). The metrics used are described in detail in the next subsection.  %

\subsection{Temporal metrics and the EVI/VOD lag} \label{metrics}

It is customary to summarize the information content of a time series using metric statistics. A variety of seasonal metrics can be used to capture crop properties such as greenup and maturity or the rate of vegetative development, providing knowledge about the ecosystem status and the crop progress. %
Optical metrics such as peak of the growing season, slope, range of measurable photosynthetic activity or growing season length are typically used as input covariates for regression  \citep{reed94, hill2003, zhong16,guan17}. Also, the use of VOD-based seasonal metrics for crop yield assessment was explored in a previous study \citep{chaparro18}. It uses the same agricultural and VOD dataset used here, but comparisons of yield and VOD metrics are performed at a 36-km pixel scale. In particular, the metrics range and standard deviation were explored as measures of the rate of growth, the metrics small and large integral were explored as accumulative proxies, and the metrics maximum and average were introduced to capture levels during the growing period. A principal component regression was then used to combine all the metrics and the first principal component (PC1) was proposed as a summarizing metric, outperforming the rest.
In this work, we explore three single-sensor metrics derived from EVI and VOD time series: (a) the maximum value, (b) the small integral (area under the growing period, see Fig \ref{fig:diagramSETUP}(b)), and (c) the PC1 metric proposed by \cite{chaparro18}. The information from start and end of season required to compute these metrics has been obtained from USDA-NASS phenology information, following the same strategy used by \cite{chaparro18}.

We additionally introduce a {\em synergistic metric} based on the time lag between EVI and VOD time series. Differences between EVI and VOD data are visible in the time domain, where a clear time-shift is observed, see Fig. \ref{fig:meanTS}. VOD peaks (in blue) are delayed with respect to EVI peaks (in red) in all counties, with time lags ranging from 6 to 79 days. This lag is directly related with the seasonal synchronicity of plant water storage (represented by VOD) and leaf phenology (represented by EVI), which significantly varies across biomes \citep{tian18}.
Over croplands, different patterns of phenology result from different crop and seed varieties. Also, different crop types respond differently to environmental stresses such as soil moisture or temperature anomalies at different stages of growth, leading to different temporal signatures. We want to emphasize here that the EVI/VOD lag is related to crop yield (Fig. \ref{fig:meanTS}). Also, note that this lag allows us to differentiate the existing crop mixture in each county, which in turn is related to the amount of production. Interestingly, counties dominated by corn lead to the shorter lags, and counties dominated by wheat to the larger lags. However, a larger sample should be studied to confirm this result. It is worth noting here that in contrast with the previous metrics, this new metric can be computed on a fixed temporal window, i.e., it is not necessary to know the start and end of seasons of the crops involved. Unlike previous metrics, the lag metric merges the temporal characteristics of the two sources in a unique value.

\begin{figure}[t!]
\centerline{
\includegraphics[height = 4cm]{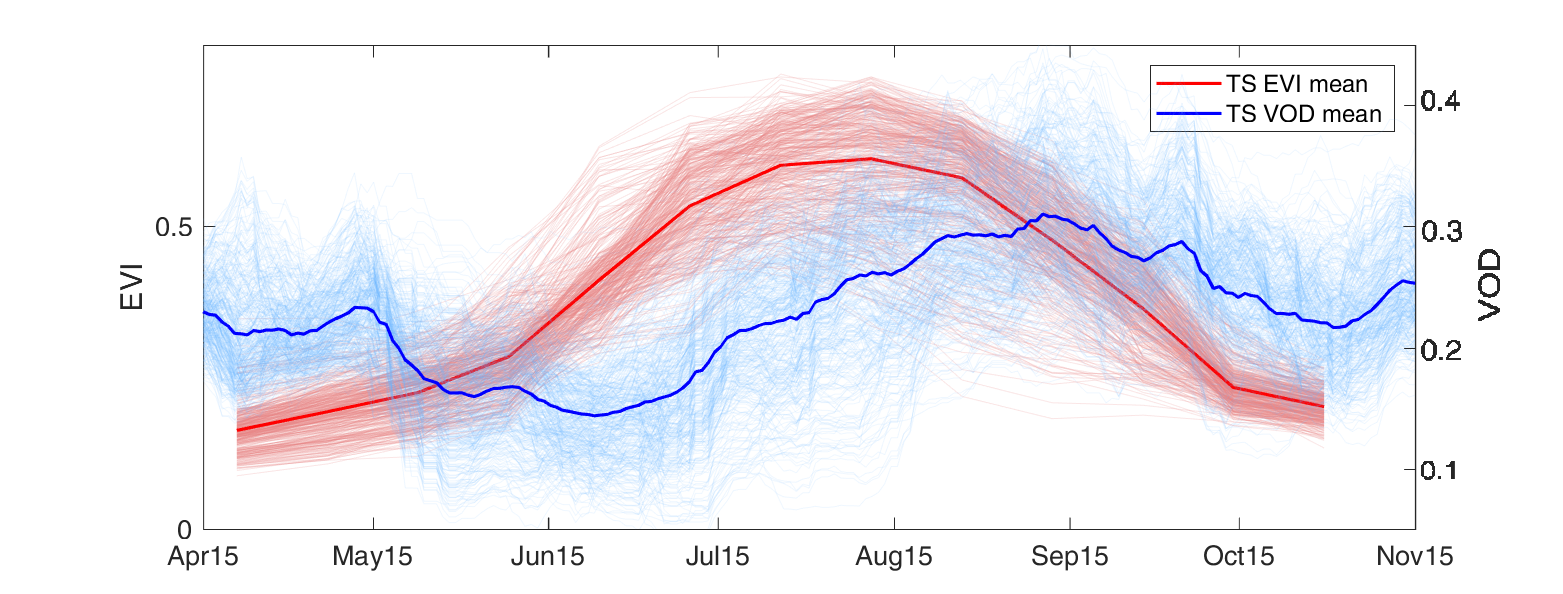}~~
	\includegraphics[height = 4cm]{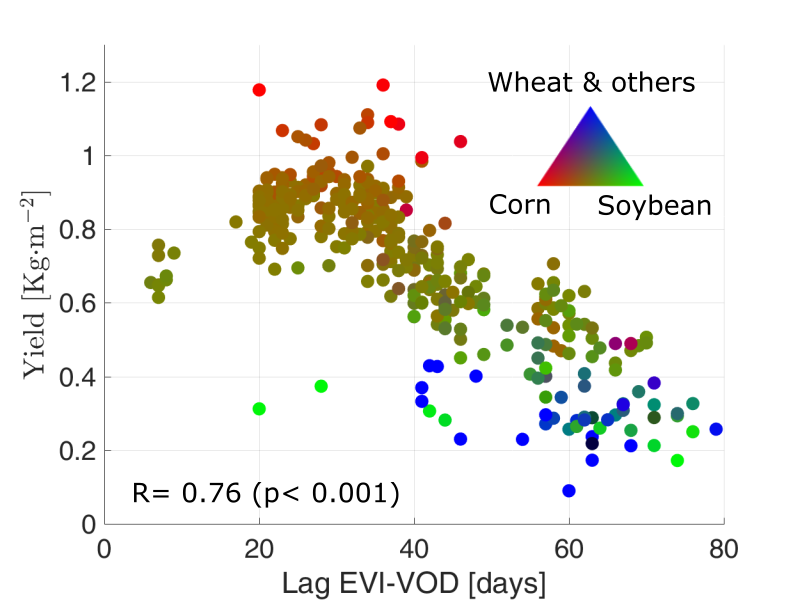}}
    \caption{Left: County-scaled time series of MODIS EVI and SMAP VOD for the study area (see Fig.1). Mean values are overlaid. Right: scatter plot showing the relationship of EVI/VOD lag and the associated yield per county. Colours represent the percentage of area planted for corn, soybean, and other crops (mainly wheat). Each vertex in the triangle corresponds to 100\%. }
    \label{fig:meanTS}
\end{figure}

\subsection{Machine learning methods} \label{ssec:ml}

In this work we also exploit metrics or time series of EVI, VOD, and their fusion, with statistical regression methods. In particular, we use a regularized linear regression (RLR) and a nonlinear (kernel-based) regression method called kernel ridge regression (KRR)~\citep{Shawetaylor04,CampsValls09wiley,Rojo17dspkm}, which has provided excellent results in previous applications involving remote sensing data over land, atmosphere and water parameter estimation~\citep{Tramontana16bg,Camps-Valls20121759,rs10050786}. The KRR method generalizes the linear regression, and generally offers improved accuracy and reduced bias.

We are given a set of $N$ county time series of $T$ observations, and the associated crop total production or crop yield in the target variable $y$. The goal is to exploit the information in $\x\in\Real^{T}$ to estimate $y\in\Real$. In multivariate (or multiple) RLR the output $y$ is assumed to be a weighted sum of $T$ observations (note that $T=1$ when using a single summarizing metric and $T=226$ when the time series of EVI and VOD are stacked together), $\x=[x_{1},\ldots,x_{T}]^\top$, that is $\hat{y}=\x^\top{\bf w} +w_o$. Maximizing the likelihood is equivalent to minimizing the sum of squared errors, and hence one can estimate the weights ${\bf w}=[w_{1},\ldots,w_{T}]^\top$ and bias $w_o$ by least squares minimization. Very often one imposes some smoothness constraints to the model and also minimizes the weights power, $\|{\bf w}\|^2$, thus leading to the RLR method.
Linear regression is however a limited approach when the relationship between the input data and the target variable is non linear. In those cases, by definition nonlinear regression can do a better job. Kernel methods constitute a family of successful methods for nonlinear regression~\citep{Shawetaylor04,CampsValls09wiley,Rojo17dspkm}. Kernel methods return a crop yield prediction for a new input vector $x_*\in\Real^T$ as:
\begin{equation}
\hat{y} = f(\x)= \sum\limits_{i=1}^{N} \alpha_i K(\x_i, \x_*) + \alpha_o,
\end{equation}
where $\{\x_i\}_{i=1}^{N}$ are county-level time series used in the training phase, $\alpha_i$ is the weight assigned to each one of them, $\alpha_o$ is the bias in the regression function, and $K$ is a kernel or covariance function (parameterized by a set of hyper-parameters) that evaluates the similarity between the test county  time series $\x_*$ and all $N$ training county time series. The process can be interpreted as a template matching: when the test time series is similar (as measured by $K$) to a training time series, a similar estimate will be given yet weighted by its relevance (as given by $\alpha_i$).
KRR essentially performs a regularized linear least squares regression in a feature space where the samples have been transformed to by a nonlinear mapping function. It can be shown that the solution is analytic, and reduces to solving a matrix inversion. As in any kernel method, one has to select the form of the similarity kernel function. In this work we selected the standard radial basis function (RBF) kernel, defined as $K(\x,\x')=\exp(-\|\x-\x'\|^2/(2\sigma^2))$, where $\sigma$ is the length-scale value. The RBF kernel is by far the most widely used in the remote sensing community because it is a smooth function able to capture the variability present in the natural world \citep{hofmann2008, theodoridis2009,Camps-Valls2009b}.

In this work, we developed RLR and KRR models for four scenarios: estimation of total crop production and estimation of corn, soybean and wheat production. In all cases, we need to choose a set of hyperparameters: the regularization $\lambda$ value in both RLR and KRR models, and the lengthscale $\sigma$ value for the KRR models. Model evaluation was done via standard cross-validation: the initial data collection was divided in two independent data sets consisting of 70\% of counties to train the models, and the remaining 30\% was reserved for the test set. All experiments were repeated $10$ times randomly selecting the training and test sets.

\section{Results and discussion} \label{sec:results}
This section presents the empirical results obtained following the proposed methodology. We first give a numerical comparison based on standard accuracy measures in the four scenarios (i.e., total, corn, soybean and wheat yield estimation) with focus on the use of metrics vs. the use of full time series, single sensor vs. multisensor, and linear vs. nonlinear regression. Then, we show the crop yield estimation maps for the best models and perform an error analysis. Next, we explore the within-season forecasting setting by shortening the adopted time window. In the last section we explore capability of the models to work with additional data in a multi-year setting. %

\subsection{Numerical \& statistical comparison}

In this subsection, we report the mean and standard deviation of the mean error (ME), root mean square error (RMSE) and coefficient of determination (R$^2$) over the $10$ different trials/repetitions performed for each general and crop-specific production model.

The results obtained for the total yield estimation models are summarized on Table \ref{tab:resultsall}. Here, we aim at estimating the total production, computed as the weighted average of all the crops planted in the region (see Section \ref{sec:data}). %
We can observe that in all experiments the nonlinear models based on KRR achieved better results than linear models based on RLR. The best model is the KRR combining EVI and VOD full time series, which explains 84\% of yield variance. Among the metrics, the proposed EVI/VOD lag outperforms the results obtained with the other metrics. It is able to explain 71\% of yield variance, whereas the second best metric (PC1) is able to explain 45\% with VOD and 38\% using EVI. Note a higher $R^2$ of 0.65 was obtained for PC1 and VOD in \citep{chaparro18}, probably due to the fact that regressions were performed at the 36 km scale, implying different season definitions and resulting metrics. The obtained results suggest that EVI and VOD time series contain complementary information that is worth exploiting jointly. Using data across optical and microwave spectral ranges enhances crop yield assessment capabilities beyond what can be achieved by using an individual sensor. We also performed experiments summarizing the time series using PCA, but results were comparable to the use of the whole time series (not shown).

It can be noted that microwave data in the form of VOD attains better results using the integral and the PC1 metrics, whereas EVI leads to better results using the maximum. This could be interpreted as microwaves being able to capture better the growth rate relation to yield, and the optical maximum being more closely related to total yield. When EVI and VOD metrics are combined in the form of PC1, better results are obtained with the microwave-based metrics. Using full time series, results with VOD are also slightly better than with EVI. The fact that the EVI product chosen is provided as a 16-day composite and VOD is provided daily could have an impact in the obtained results. Weekly composites of EVI are expected to lead to a better characterization of the optical signature and should be explored in further studies. Also, advanced distribution regression methods could be explored to further exploit the native spatio-temporal information on the satellite data in the models \citep{adsuara19}.
Still, our results show that systematically, the combination of the optical and microwave data in the form of a single metric (the EVI/VOD lag) or stacking the time series leads to the best overall results. Particularly, the use of full time series improved the results significantly and made the processing simpler (no need of crop season definition or metric computation) and more straightforward.

The results for the crop-specific experiments (corn, soybean and wheat) for the proposed EVI/VOD lag metric and for the three full time series configurations are shown on Table \ref{tab:resultscrops}. This is a more challenging scenario, since crop-type information at the parcel or sub-county level is not available, and the averaged satellite time series at the county scale contain aggregated information from mixed crops. The best model for crop-specific yield estimation is also the KRR combining EVI and VOD full time series. A highest performance was obtained for both corn and soybean with an $R^2\geq 0.9$, and a lower goodness-of-fit for wheat ($R^2=0.72$). The EVI/VOD lag metric is able to explain the 73\% and 60\% of corn and soybean yield variance, respectively, but fails at estimating wheat. The capacity of the proposed synergistic metric to estimate corn and soybean production is remarkable, since single-sensor metrics over the same region failed at performing crop-specific estimates in previous approaches \citep{chaparro18}. The fact that corn and soybean are the most abundant crops (in both extent and production) in the study area might explain the good results obtained for these crops. Wheat, on the contrary, is not well represented in the area and hence in the dataset. It only predominates across North Dakota and northwestern Minnessota, being present in a very low proportion in the rest of the study area (see Fig.\ref{fig:totalhist}). We hypothesize this directly translates in worse results. Note, however, that even in the more challenging scenario of wheat crop yield estimation, the synergistic use of optical and passive microwave data boosts the results. %

\begin{table}[t] \centering %
\caption{Results for the estimation of total yield: RMSE (x100) [kg$\cdot$m$^{-2}$] and R$^2$.}
\resizebox{\textwidth}{!} {
\begin{tabular}{ll|c|c|c|c}
\hline \hline
\multicolumn{1}{c}{Model} &\multicolumn{1}{c}{$T$}  & \multicolumn{2}{|c}{RLR} & \multicolumn{2}{|c}{KRR} \\
\hline
$N=385$ counties &  & {RMSE} & {R$^2$} & {RMSE} & {R$^2$}  \\
\hline\hline
VOD$_{max}$ & 1&20.22 $\pm$ 0.62&0.15 $\pm$ 0.04&20.53 $\pm$ 0.83&0.13 $\pm$ 0.05\\
EVI$_{max}$ & 1&17.68 $\pm$ 0.8&0.34 $\pm$ 0.06&53.73 $\pm$ 113.47&0.31 $\pm$ 0.14\\
VOD$_{int}$ & 1&19.52 $\pm$ 0.85&0.22 $\pm$ 0.05&16.91 $\pm$ 0.7&0.41 $\pm$ 0.05\\
EVI$_{int}$ & 1&21.53 $\pm$ 1.07&0.02 $\pm$ 0.02&45.26 $\pm$ 53.17&0.09 $\pm$ 0.05\\
VOD$_{pc1}$  & 1 &17.72 $\pm$ 1.19&0.34 $\pm$ 0.02&16.21 $\pm$ 1.55&0.45 $\pm$ 0.07\\
EVI$_{pc1}$ & 1&18.04 $\pm$ 1.23&0.36 $\pm$ 0.05&17.67 $\pm$ 1.47&0.38 $\pm$ 0.07\\
EVI/VOD$_{lag}$ & 1 &13.51 $\pm$ 0.96&0.61 $\pm$ 0.05& \bf{11.92 $\pm$ 1.46 }& \bf{0.71 $\pm$ 0.04}\\

\hline
\hline
EVI & 13 &10.97 $\pm$ 0.95&0.74 $\pm$ 0.03&9.8 $\pm$ 1.16&0.79 $\pm$ 0.04\\
VOD & 213 &9.51 $\pm$ 1.33&0.79 $\pm$ 0.06&9.23 $\pm$ 1.23&0.81 $\pm$ 0.05\\
EVI+VOD &  226 &8.85 $\pm$ 0.92&0.84 $\pm$ 0.03& \bf{8.72 $\pm$ 0.91}& \bf{0.84 $\pm$ 0.03}\\
\hline
\hline

\end{tabular}}
\label{tab:resultsall}
\end{table}

\begin{table}[t!] \centering %
\caption{Results for the estimation of corn, soybean and wheat yields: RMSE (x100) [kg$\cdot$ m$^{-2}$] and R$^2$.}
\resizebox{\textwidth}{!} {
\begin{tabular}{ll|c|c|c|c}
\hline \hline
\multicolumn{1}{c}{Model} &\multicolumn{1}{c}{$T$}  & \multicolumn{2}{|c}{RLR} & \multicolumn{2}{|c}{KRR} \\
\hline
&  & {RMSE} & {R$^2$} & {RMSE} & {R$^2$}  \\
\hline \hline
{Corn, $N=363$ counties} \\
\hline
EVI/VOD$_{lag}$ & 1 &10.45 $\pm$ 0.56&0.69 $\pm$ 0.03&9.79 $\pm$ 0.28&0.73 $\pm$ 0.02\\
EVI & 13 & 9.17 $\pm$ 0.88&0.76 $\pm$ 0.04&7.77 $\pm$ 1.03&0.83 $\pm$ 0.03\\
VOD & 213 &7.21 $\pm$ 0.54&0.85 $\pm$ 0.02&6.72 $\pm$ 0.67&0.87 $\pm$ 0.02\\
EVI+VOD & 226 &5.95 $\pm$ 0.6&0.9 $\pm$ 0.03&  \bf{5.6 $\pm$ 0.51} & \bf{0.91 $\pm$ 0.02}\\
\hline\hline

{Soybean, $N=361$ counties} \\
\hline
EVI/VOD$_{lag}$ & 1 & 4.3 $\pm$ 0.26&0.51 $\pm$ 0.04&3.92 $\pm$ 0.23&0.6 $\pm$ 0.03\\
EVI & 13 &2.68 $\pm$ 0.18&0.82 $\pm$ 0.04&2.2 $\pm$ 0.18&0.88 $\pm$ 0.03\\
VOD & 213 &2.38 $\pm$ 0.18&0.85 $\pm$ 0.04&2.3 $\pm$ 0.18&0.86 $\pm$ 0.04\\
EVI+VOD& 226 &  \bf{2.06 $\pm$ 0.23} & \bf{0.9 $\pm$ 0.03} &  \bf{2.05 $\pm$ 0.24} & \bf{0.9 $\pm$ 0.03}\\
\hline\hline

{Wheat, $N=204$ counties} \\
\hline
EVI/VOD$_{lag}$ & 1 &9.14 $\pm$ 0.73&0.03 $\pm$ 0.04&9.18 $\pm$ 1.21&0.06 $\pm$ 0.05\\
EVI& 13 &7.35 $\pm$ 0.72&0.31 $\pm$ 0.08&6.74 $\pm$ 0.83&0.42 $\pm$ 0.13\\
VOD& 213 &5.8 $\pm$ 0.41&0.59 $\pm$ 0.11&5.38 $\pm$ 0.52&0.64 $\pm$ 0.12\\
EVI+VOD & 226 &5.55 $\pm$ 0.45&0.65 $\pm$ 0.06&  \bf{4.96 $\pm$ 0.46} &\bf{0.72 $\pm$ 0.05}\\
\hline\hline\\
\end{tabular}}
\label{tab:resultscrops}
\end{table}

\subsection{Crop yield prediction maps and error assessment}

In the previous subsection, we trained and cross-validated the models. The model leading to the best results in all the experiments was the KRR method with EVI and VOD time series, c.f. highlighted in bold in Tables \ref{tab:resultsall} and \ref{tab:resultscrops}. Here, we apply this model to the whole study area to generate spatially explicit crop yield maps.

\begin{figure}[t!]
\begin{center}
\setlength{\tabcolsep}{-1mm}
\begin{tabular}{ccc}
\hspace{-1cm}
\bf{Survey data [kg$\cdot$m$^{-2}$]} & \bf{KRR prediction [kg$\cdot$m$^{-2}$]} &  \bf{ $\varepsilon_r$ [\%] } \\
\vspace{-0.75cm}
\hspace{-1cm}
\includegraphics[scale=0.4]{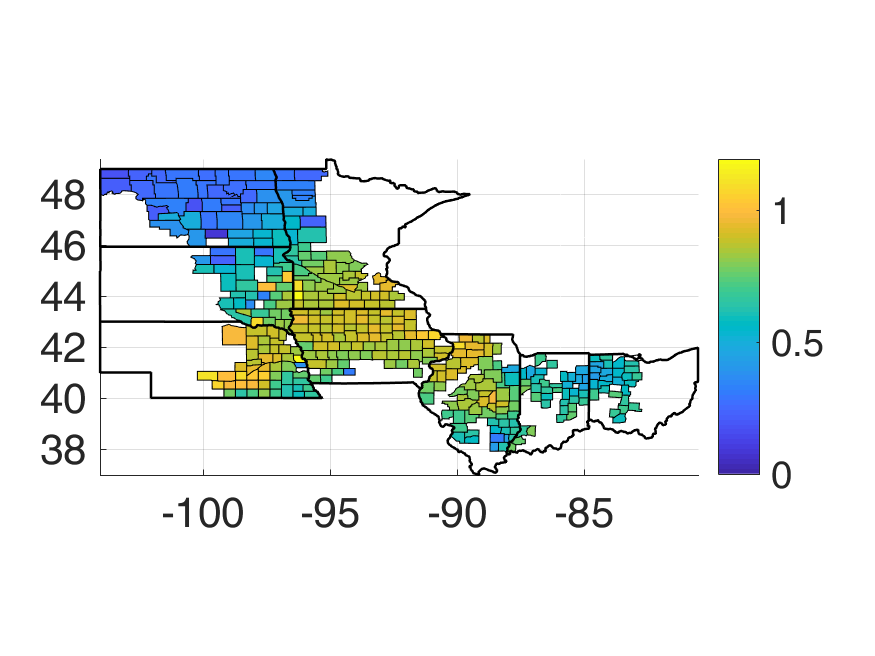} &
\includegraphics[scale=0.4]{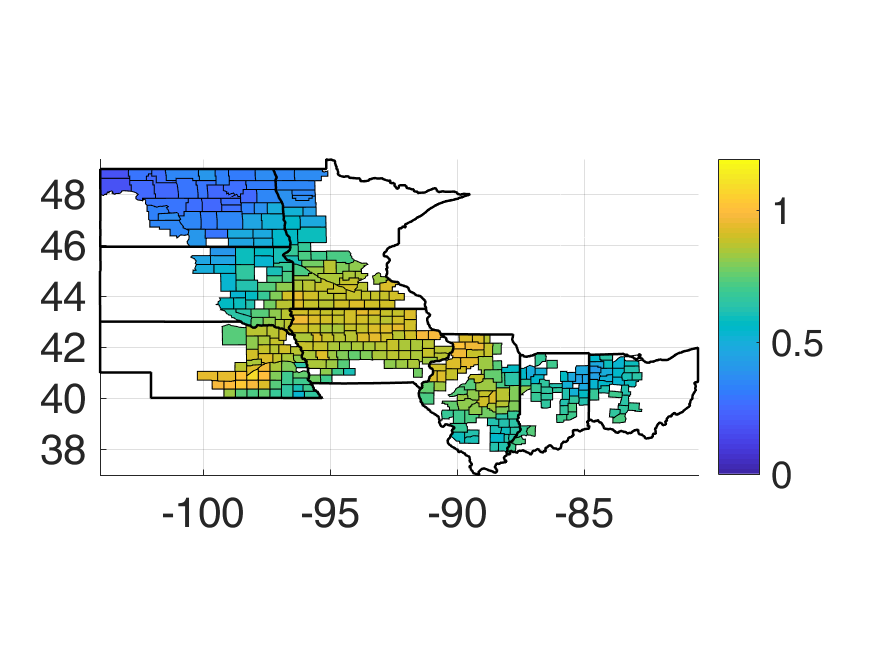}&
\includegraphics[scale=0.4]{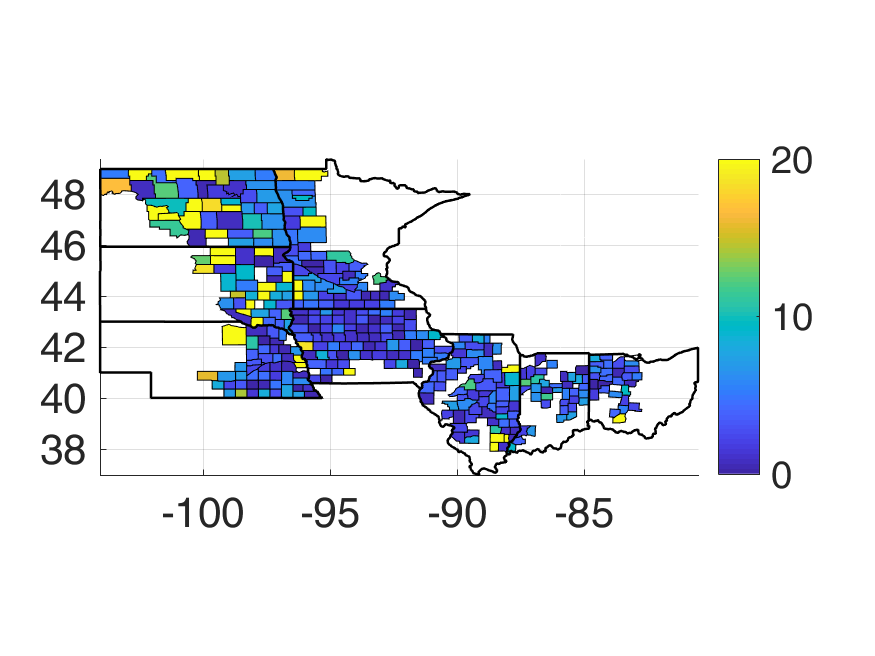}\\
\vspace{-0.75cm}
\hspace{-1cm}
\includegraphics[scale=0.4]{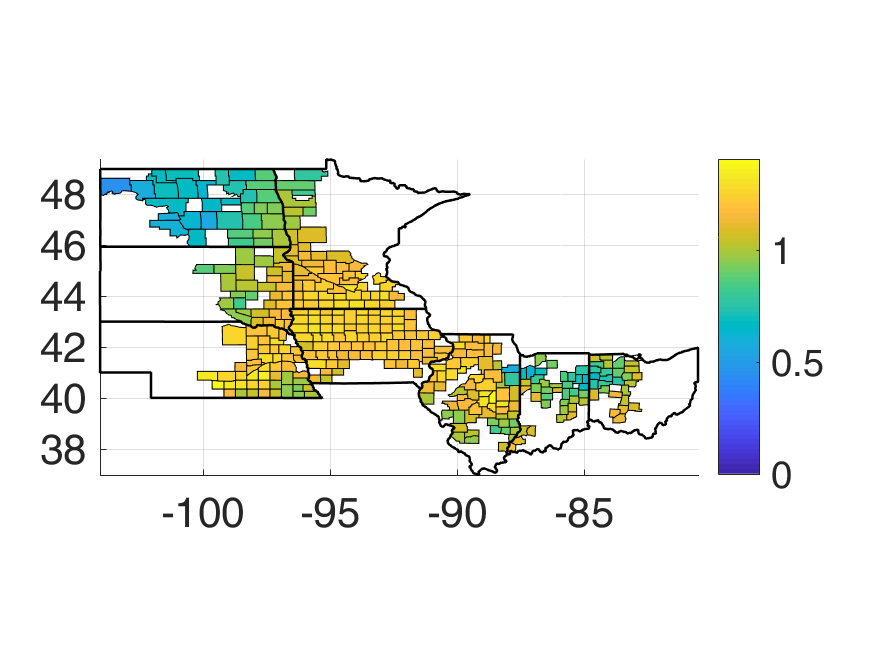} &
\includegraphics[scale=0.4]{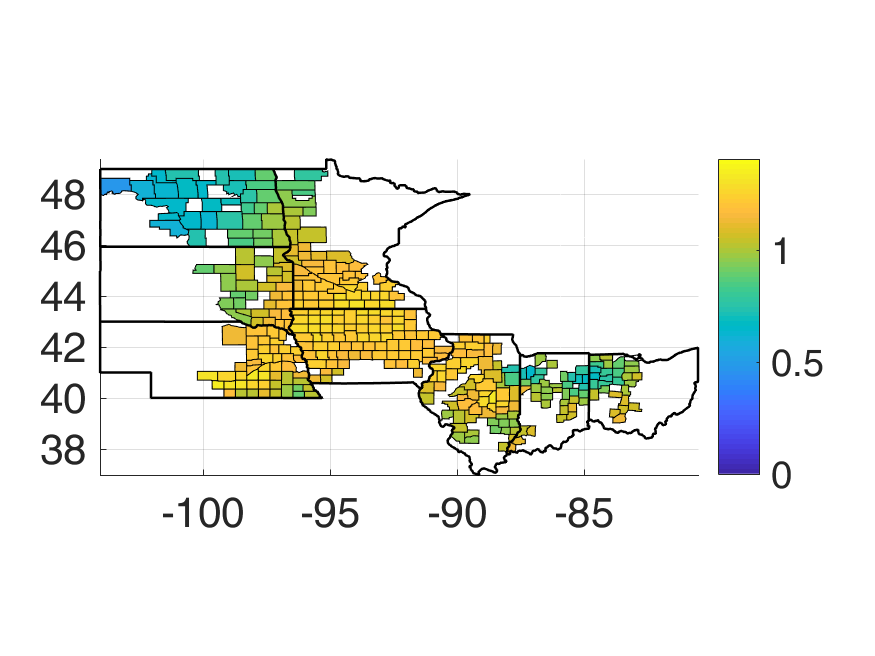}&
\includegraphics[scale=0.4]{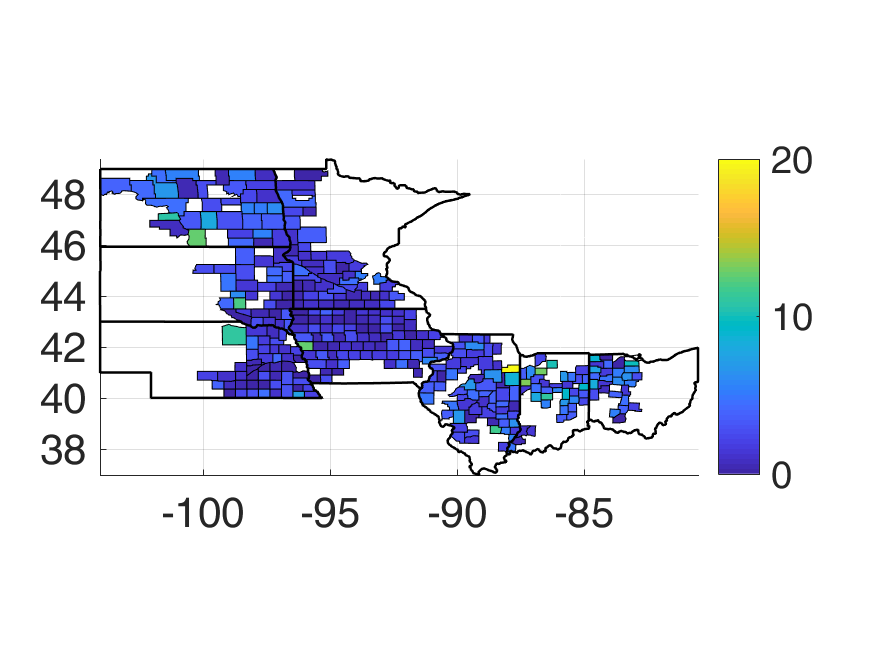}\\

\vspace{-0.75cm}
\hspace{-1cm}
\includegraphics[scale=0.4]{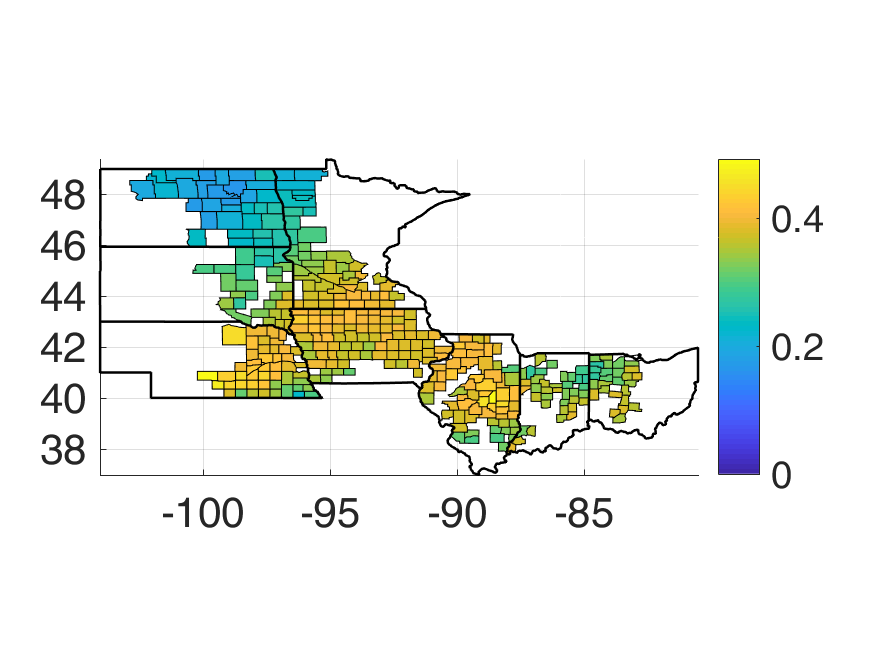} & \includegraphics[scale=0.4]{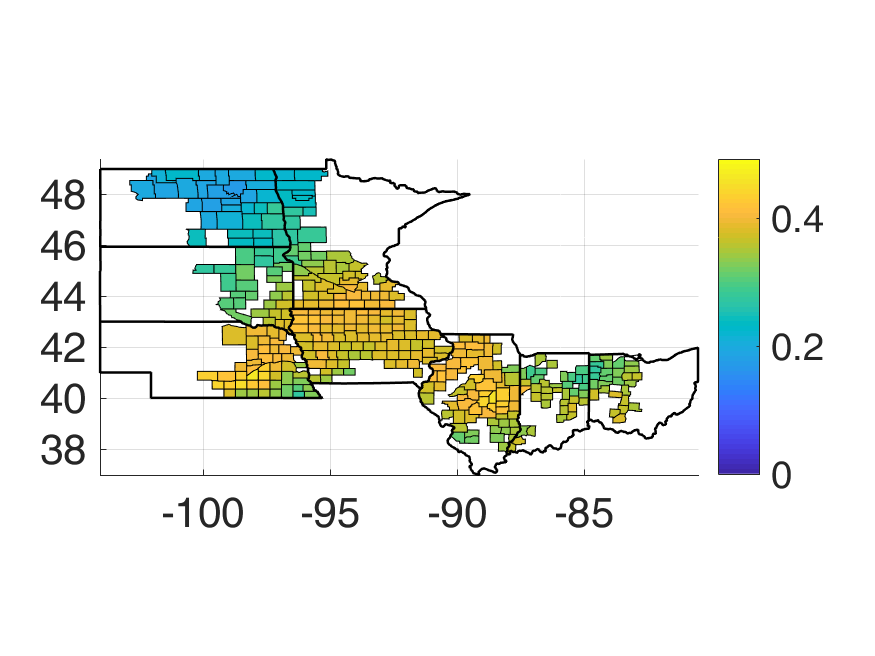}&
\includegraphics[scale=0.4]{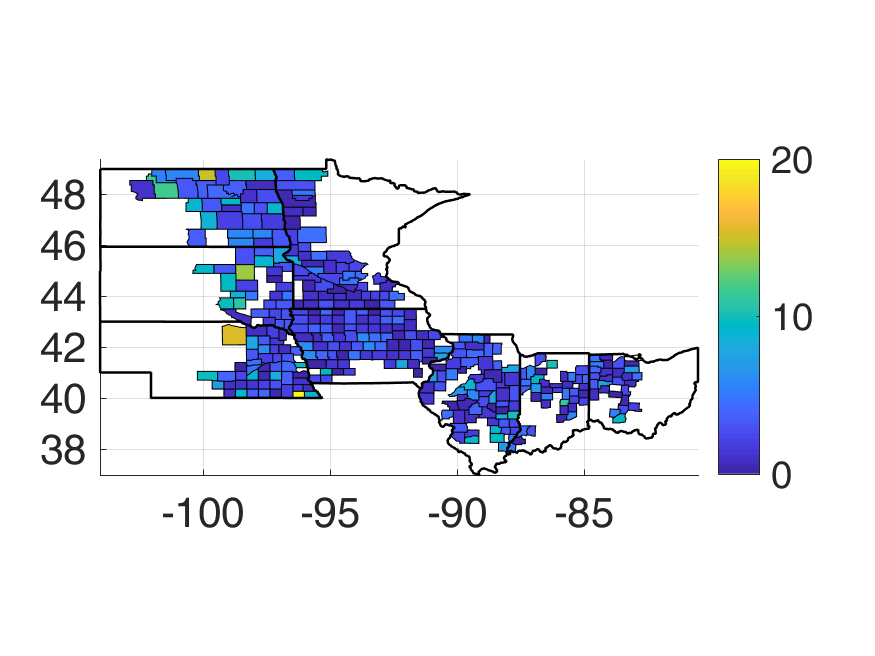}
\\
\vspace{-0.75cm}
\hspace{-1cm}
\includegraphics[scale=0.4]{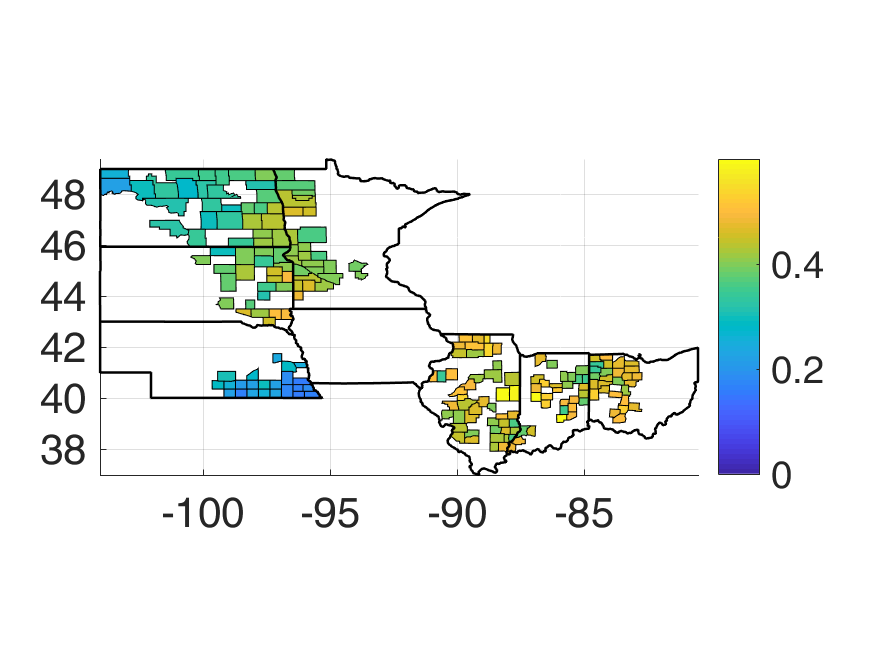} &
\includegraphics[scale=0.4]{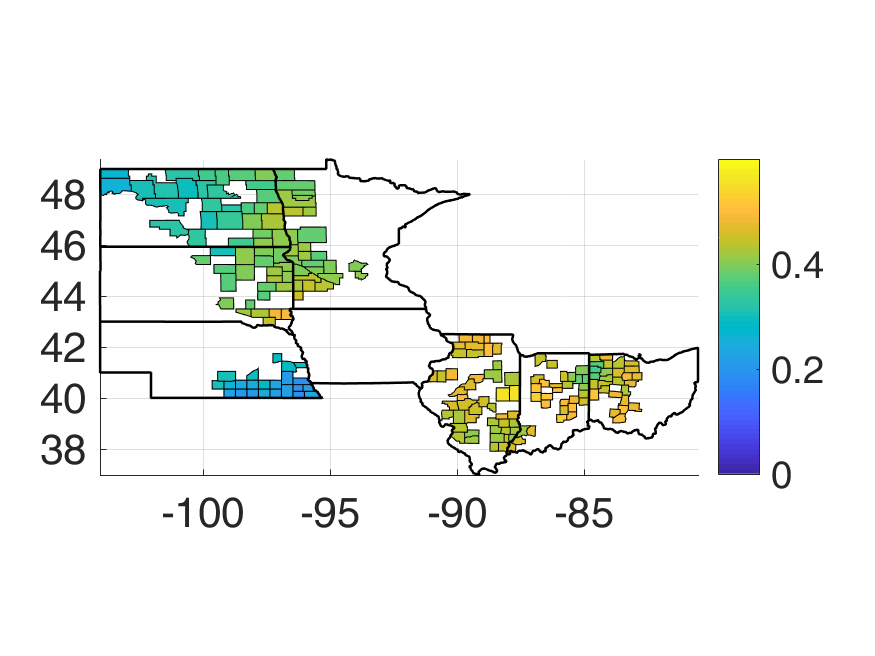}&
\includegraphics[scale=0.4]{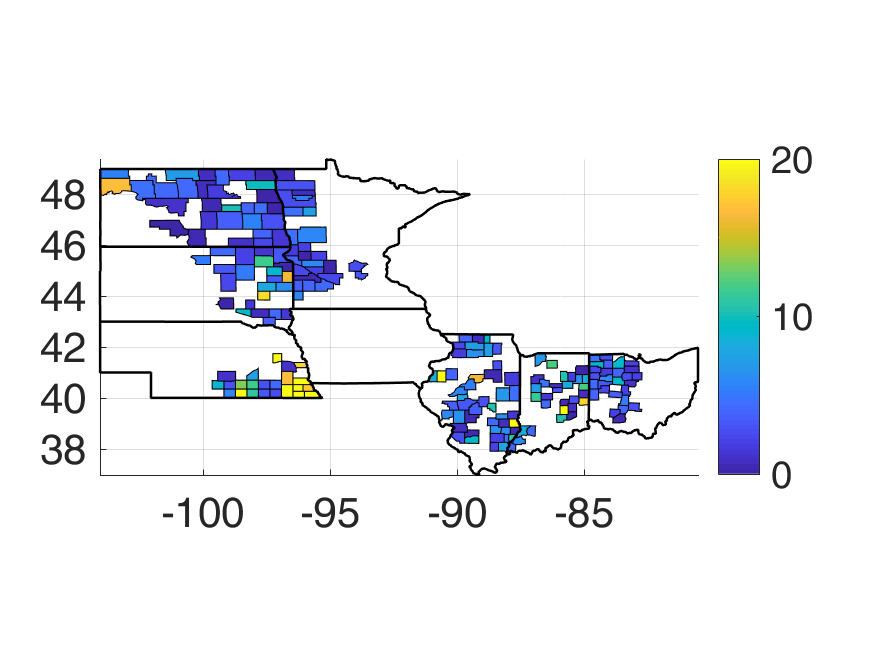}
\\
\end{tabular}
\end{center}
\caption{County-level maps of yield survey data [$kg/m^2$] (left), best yield estimation obtained with full time series and KRR [$kg/m^2$] (in bold text on Tables \ref{tab:resultsall} and \ref{tab:resultscrops}) (middle) and relative error [\%] (right). From top to bottom: total yield, corn, soybean and wheat crop yields.}
\label{fig:mapsall}
\end{figure}

Figure \ref{fig:mapsall} shows column-wise the yield survey data provided by USDA-NASS (taken here as the ground truth for training and validation), the crop yield estimation obtained with the KRR method on EVI+VOD time series, and the relative error maps between the survey data and the estimations.
First row shows the maps for the total crop yield experiment (Table \ref{tab:resultsall}). The higher reported production is located in central latitudes in yellowish colours, and the lowest in the north and northwest. The crop yield model estimates a very similar spatial distribution, with relative errors lower than 20\% in all counties. The worst results are located in the northwest of the study region, where the production of total crop yield is lower and wheat predominates (see Fig. \ref{fig:totalhist}). Note also that counties in this region are ``atypical'' in terms of crop types, since the corn-soybean mixture largely predominates. These two issues (quantity and diversity) are challenging for statistical algorithms working at such coarse spatial resolutions.

Second to fourth rows in Fig. \ref{fig:mapsall} show the maps for the three crop-specific estimates: corn, soybean and wheat. We can see the similarity between the spatial patterns between corn and soybean yield. The two crops coexist across the study area, although their cultivation is more extensive in central latitudes. Wheat production, in turn, is distributed unevenly across the study area, and 181 counties out of 385 do not cultivate it (e.g., Iowa). The good results shown previously in Tables \ref{tab:resultsall} and \ref{tab:resultscrops} with low RMSE values and high R$^2$ values are confirmed here where a high proportion of counties show relative errors lower than 10\% in all crop types.

\begin{figure}[t!]
\setlength{\tabcolsep}{0mm}
\begin{tabular}{cc}
    {\bf Total yield} & {\bf Corn}  \\
	\includegraphics[height=45mm]{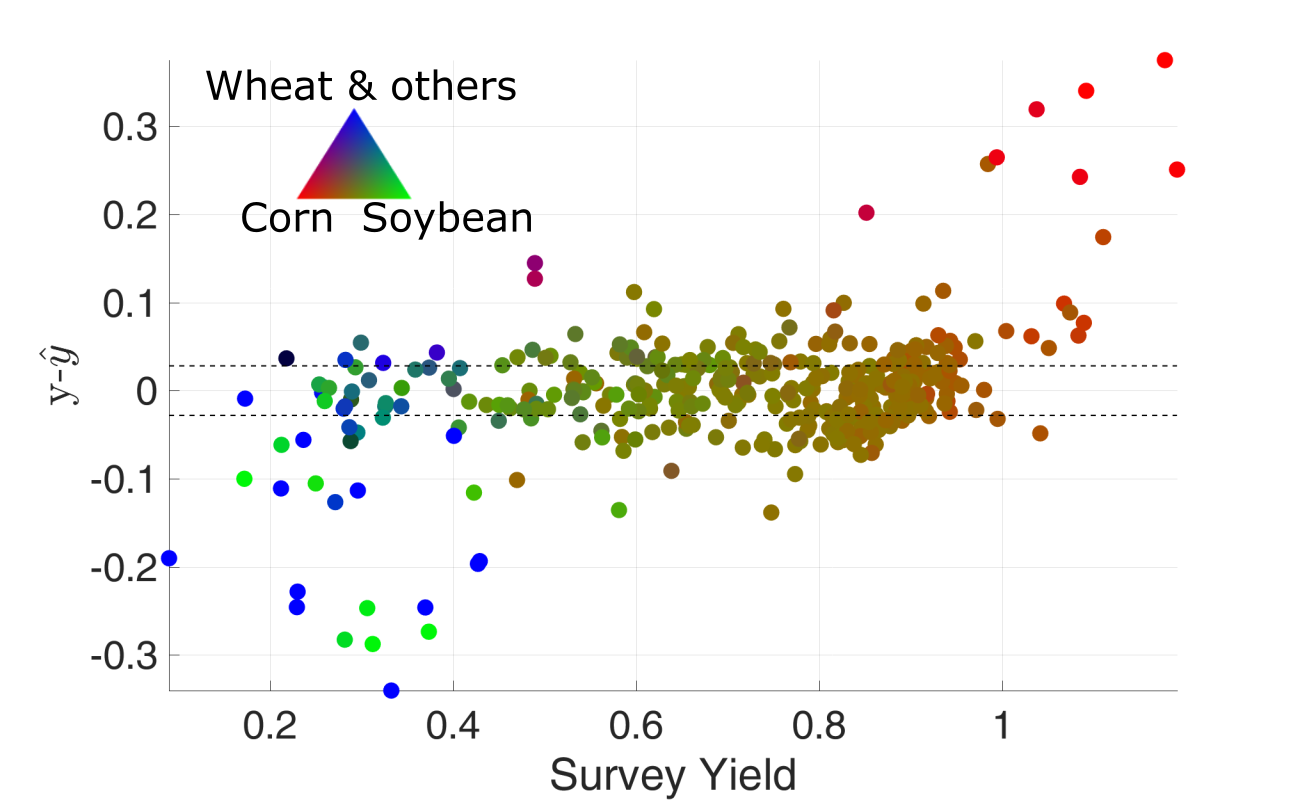} &
	\includegraphics[height=45mm]{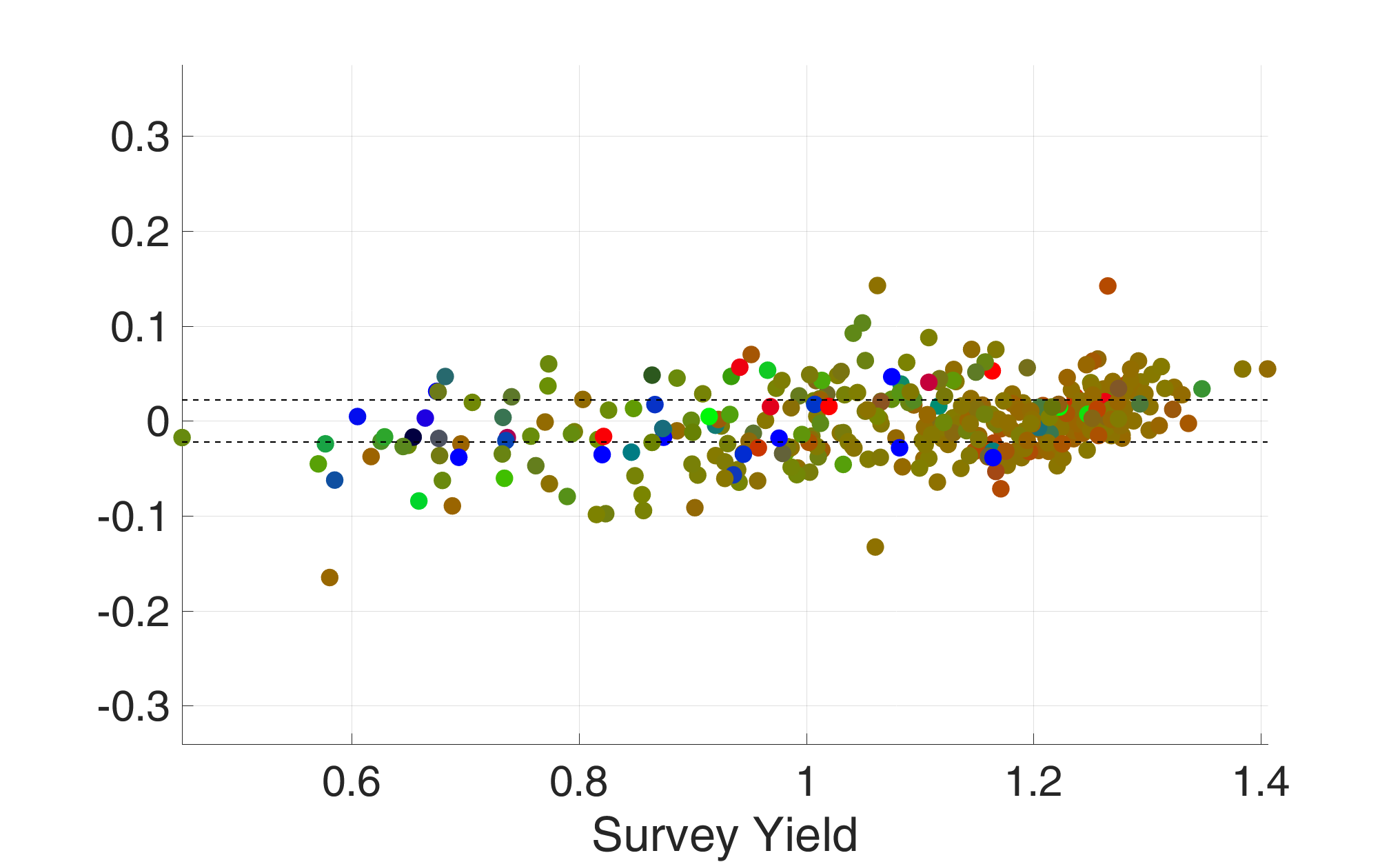}\vspace{0.5 cm}\\
	{\bf Soybean} & {\bf Wheat}  \\
	\includegraphics[height=45mm]{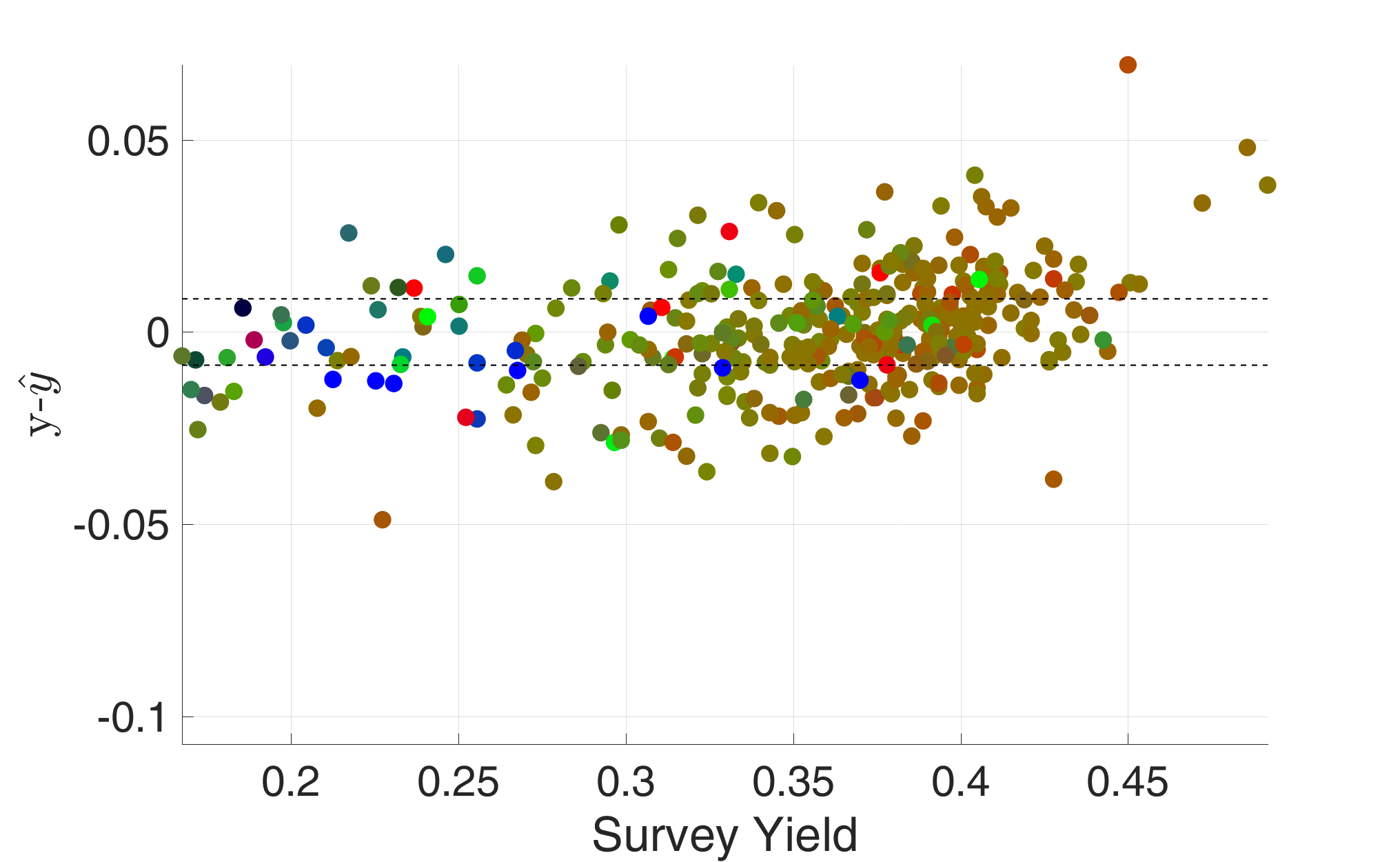} &
	\includegraphics[height=45mm]{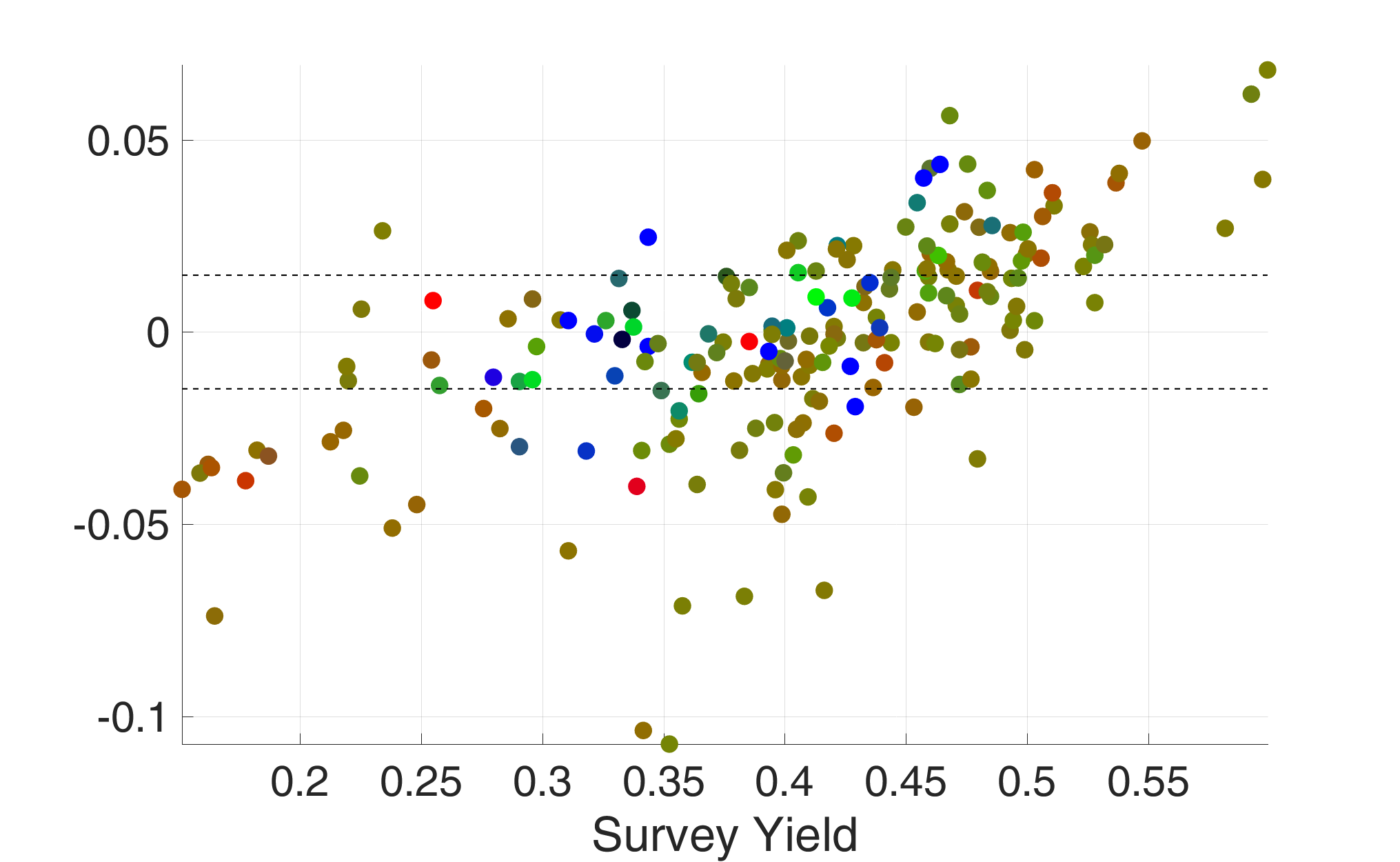}\\
	\end{tabular}
    \caption{Scatter plots showing the residuals in the estimation of crop yield ($y-\hat{y}$) [kg$\cdot$m$^{-2}$] vs. the survey yield for the best prediction model and the four sceanrios: total yield, corn, soybean and wheat yield (in bold, from Tables \ref{tab:resultsall} and \ref{tab:resultscrops}). Colors depict the percentage of corn, soybean and wheat planted within each county. Dashed lines indicate the 25th-75th percentiles of the residuals distribution.}
    \label{fig:modeall}
\end{figure}

Despite the low errors obtained overall, it is important to make further assessments to identify their possible causes. Understanding the behaviour of the proposed model for crop yield estimation is key for its extension and application in multi-annual estimates and diverse agro-ecosystems. %
Figure \ref{fig:modeall} shows the residuals in the estimation of crop yield as a function of the survey yield obtained for the best prediction model in each scenario: total yield, corn, soybean, and wheat yield.
For the total yield experiments, it can be seen that the counties with poorest predictions contain a single variety of crop (i.e., homogeneous, colours from each vertex of the triangle). In particular, the model underestimates counties close to 100\% corn (positive residuals) and 100\% soybean and overestimates counties closed to 100\% wheat (negative residuals). This result can be explained by the fact that the model has been trained in counties where mostly corn and soybean were farmed (see Fig. \ref{fig:totalhist}), and hence best results are obtained for the counties with the most ``typical" crop mixture. The errors of the proposed crop yield estimation model therefore depend on the coexistence of different crops in each county and on how representative the mixture of crops is within the study area as a whole.
Regarding crop-specific experiments, the residuals of corn and soybean models are distributed uniformly across all yield values. These two models estimate correctly the yield regardless of the mixture of crops in each county. This is probably due to the fact that counties with 100\% wheat production are not included. Finally, the residuals of the wheat experiment seem to depend on survey yield itself. Counties with survey yield values lower than 0.45 kg$\cdot$m$^{-2}$ are likely to be underestimated, contrary to the case of counties with higher survey yield. %

\begin{figure}[t!]
\setlength{\tabcolsep}{0mm}
\begin{tabular}{ccccc}
    & {\bf Total yield} & {\bf Corn}  & {\bf Soybean} & {\bf Wheat}\\
    \includegraphics[scale = 0.55]{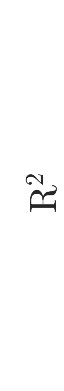}&
    \includegraphics[scale = 0.65]{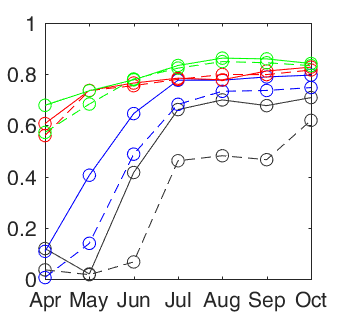}& \includegraphics[scale = 0.65]{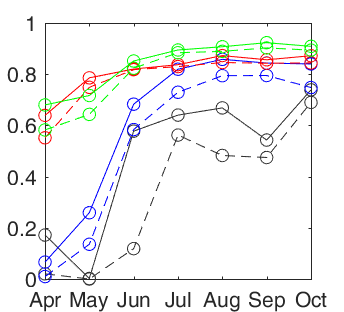}& \includegraphics[scale = 0.65]{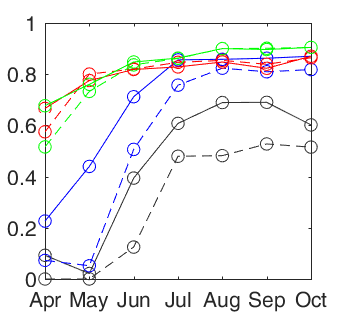}&
    \includegraphics[scale = 0.65]{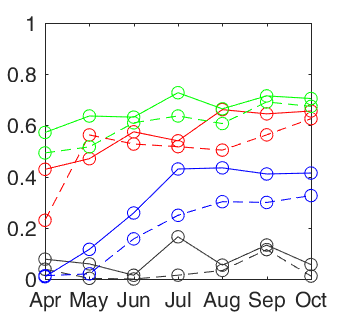}\\
    \includegraphics[scale = 0.55]{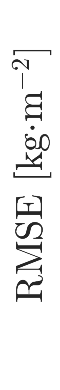}&
    \includegraphics[scale = 0.65]{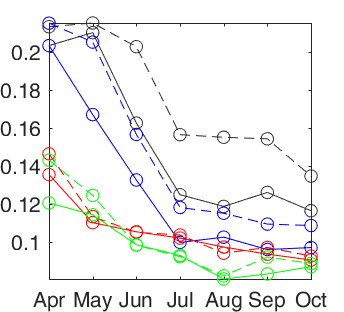}& \includegraphics[scale = 0.65]{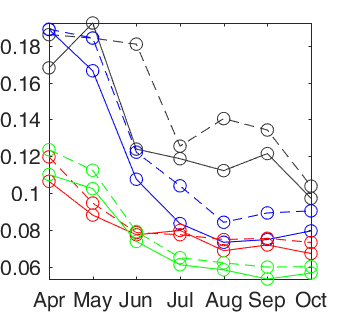}& \includegraphics[scale = 0.65]{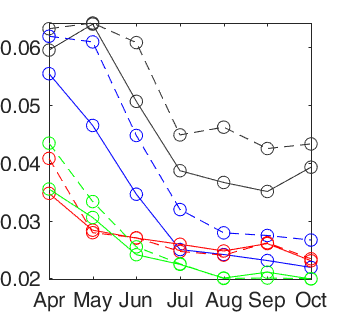}&
    \includegraphics[scale = 0.65]{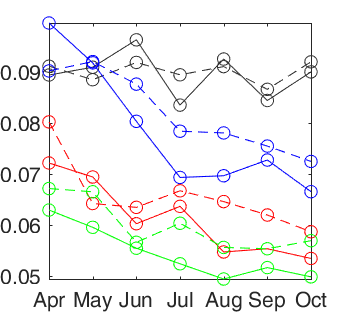}\\
    \end{tabular}
    \begin{tabular}{r}
     \includegraphics[scale = 0.4]{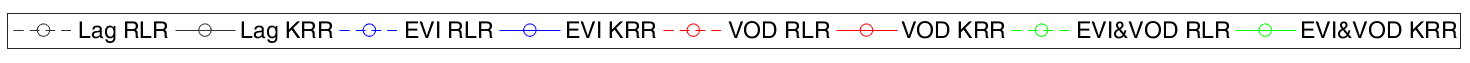}     \\
    \end{tabular}
    \caption{R$^2$ and RMSE [kg$\cdot$ m$^{-2}$] evolution as a function of the temporal window used to develop the models. }
    \label{fig:evolution}
\end{figure}

\subsection{Limits of yield prediction}

The ability of a model to make within-season predictions poses a more challenging problem than the estimation of crop yield right before harvesting, since a more limited amount of input information is available to train it.  %
In this subsection we develop models with shorter time windows and analyze their performance for crop yield prediction. %
Figure \ref{fig:evolution} compares R$^2$ and RMSE values for total yield and crop-specific yield predictions using different time windows from 1 to 7 months with monthly increments. We used our proposed EVI/VOD lag metric as well as the full time series using both RLR and KRR models with EVI, VOD and EVI+VOD. For each experimental setup, we trained the models and cross-validated them as in the previous sections. %
As expected, lowest prediction performances are obtained with the EVI/VOD lag metric, since it requires a minimum number of months to achieve a good performance. In any case, the metric is able to predict total yield, corn and soybean yield four months before harvest with $R^2>0.5$, yet it fails to predict wheat yield ($R^2<0.2$). %
Using time series, the synergy of sensors always leads to the best predictions, especially when using the nonlinear (KRR) methods. From July onward, results reach a maximum plateau in accuracy (RMSE reduces approximately by a half and R$^2$ is close to the maximum), which suggests that the best models (KRR and EVI+VOD time series) can achieve reasonably accurate crop predictions four months in advance. Better predictions are obtained with VOD-only than with EVI-only time series, specially at the shorter time windows where the EVI $16$-day composite is clearly a limitation. Also, the differences in performance between EVI and VOD-only time series are the largest for wheat prediction, where information from microwaves seem to explain significantly better the yield variability. Further results with a larger sample are nevertheless needed to confirm the crop-type prediction capabilities of the proposed models.

\subsection{Multi-year setting} \label{multiyear}

\begin{figure}[t!]
\setlength{\tabcolsep}{0mm}
\begin{tabular}{c}
     {\bf Inputs}  \\
     \includegraphics[scale = 0.25 ]{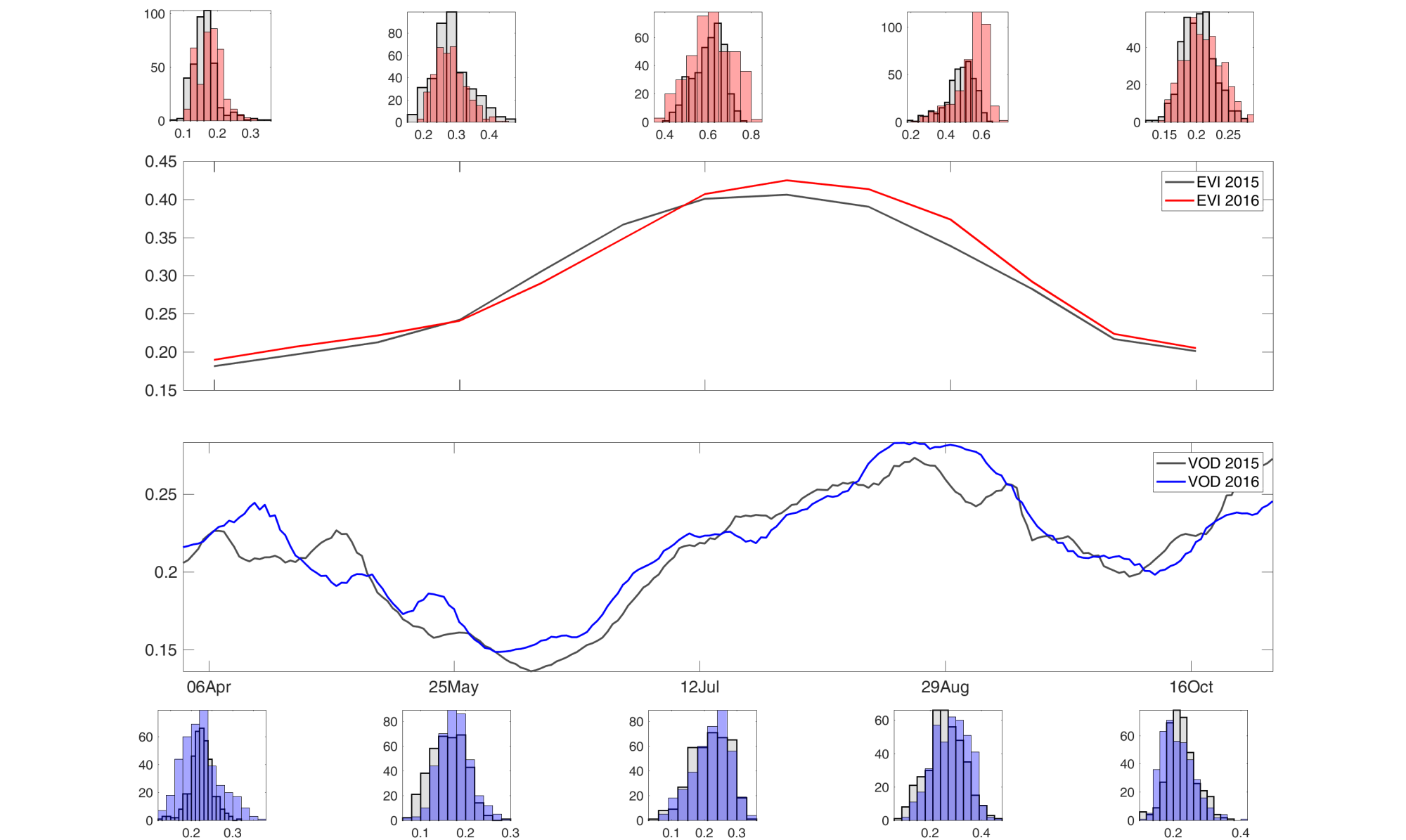}\\
\end{tabular}
\begin{tabular}{ccc}
    {} & {\bf Outputs} & {}\\

	\includegraphics[height = 3.5cm]{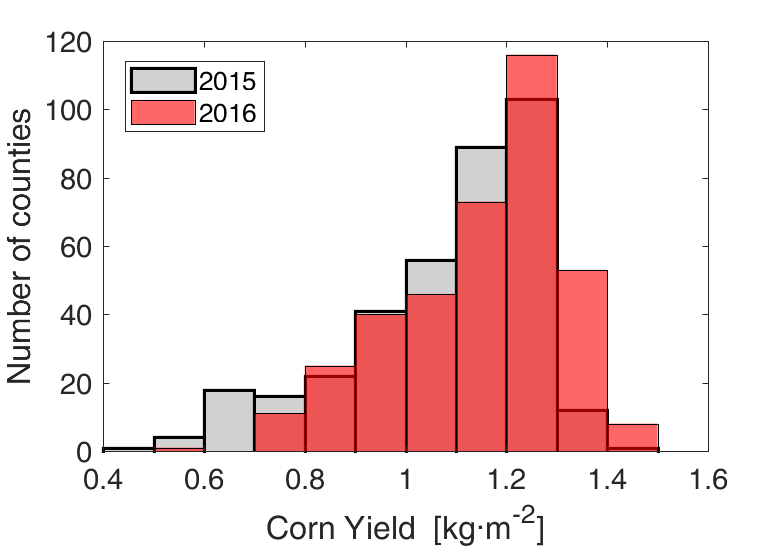} &
	\includegraphics[height = 3.5cm]{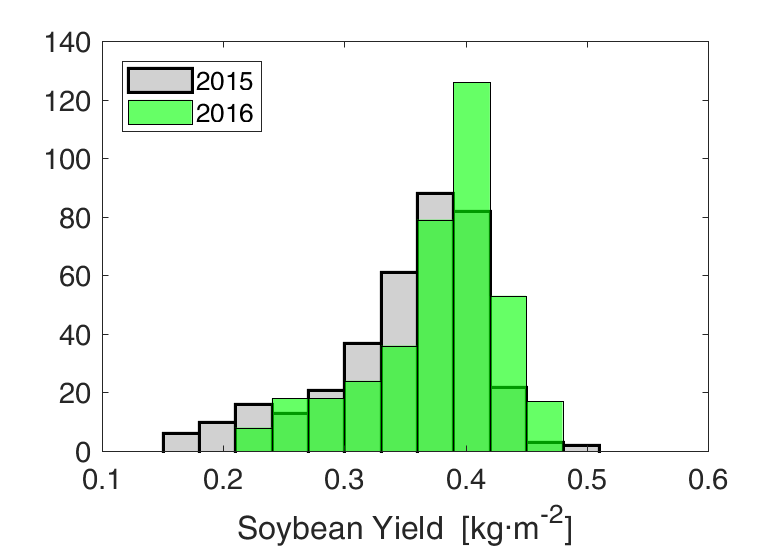} &
	\includegraphics[height = 3.5cm]{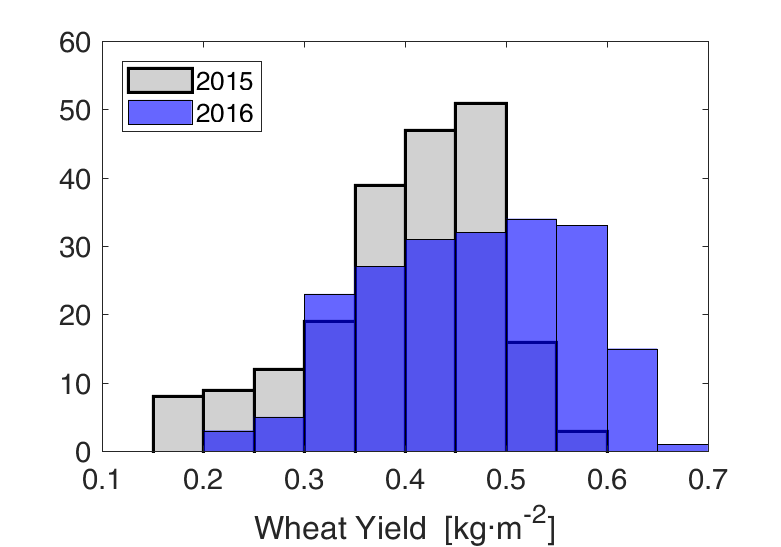}	\\
	\end{tabular}
    \caption{\blue{Top: mean time series of EVI and VOD data of both 2015 and 2016 years. Histograms plot showing the input values for specific time. Bottom: Histograms on the reported yield per county for crops scenarios with 2015 data and 2016 data.}}
    \label{fig:histograms2016}
\end{figure}

\begin{table}[t!] \centering %
\caption{Results for the estimation of all crops, corn, soybean and wheat yields with two years of data: RMSE (x100) [kg$\cdot$ m$^{-2}$] and R$^2$.}
\resizebox{\textwidth}{!} {
\begin{tabular}{ll|c|c|c|c}
\hline \hline
\multicolumn {1}{c}{Model} & \multicolumn{1}{c}{$N$}  & \multicolumn{2}{|c}{RLR} & \multicolumn{2}{|c}{KRR} \\
\hline
$T = 226$ &&   {RMSE} & {R$^2$}  & {RMSE} & {R$^2$}  \\
\hline \hline
EVI+VOD All & 776 & 9.24 $\pm$ 0.55 & 0.8 $\pm$ 0.03 & 8.76 $\pm$ 0.6 & 0.82 $\pm$ 0.03\\
EVI+VOD Corn & 736 & 8.18 $\pm$ 0.84 & 0.8 $\pm$ 0.04 & 6.31 $\pm$ 0.35 & 0.88 $\pm$ 0.01\\
EVI+VOD Soybean & 740 &  2.54 $\pm$ 0.24 & 0.83 $\pm$ 0.04 & 1.9 $\pm$ 0.12 & 0.91 $\pm$ 0.02\\
EVI+VOD Wheat  & 408 &6.15 $\pm$ 0.33 & 0.61 $\pm$ 0.05 & 5.52 $\pm$ 0.27 & 0.69 $\pm$ 0.04\\
\hline\hline
\end{tabular}}
\label{tab:response_estimation}
\end{table}

In the previous subsections, we showed the performance of the developed models using satellite and survey data from a single year. Here, we apply the best performing models (i.e., the RLR and the KRR methods with full EVI and VOD time series) in a multi-year setting, using data from years 2015 and 2016. \blue{Firstly, we} applied the same methodology to pre-process the data and develop the new models (see Sec. \ref{sec:methods}). Note that in this new setting, the number of counties $N$ is almost doubled in all scenarios, while the number of features $T$ is maintained. This in principle will lead to a better conditioning of the problem since the ratio $N/T$ increases. %

The difference in \blue{input values and} production in the study region between years 2015 and 2016 is illustrated in Figure~\ref{fig:histograms2016}: Higher corn, soybean and wheat yields were obtained in 2016 with respect to 2015. The crop proportion within the counties remained almost invariant (not shown). Results for the estimation of crop yield with two years of data in all scenarios are shown in Table \ref{tab:response_estimation}. Several conclusions can be derived. Firstly, similar values of RMSE and R$^2$ are obtained when using one or two years of data. Secondly, best estimates are obtained for corn and soybean, followed by all crops and wheat. As expected, the nonlinear regression method is able to better adapt to the new setup and the improvement of KRR with respect to RLR is more evident than in the single year setting. Results from this section confirm the capability of the proposed models to fit additional data, and in turn, that the cross-validation procedure was a robust strategy for model development, given that the curse of dimensionality is actually reduced by keeping the ratio $N/T$ higher.

\blue{Lastly, despite not being the aim of this work, we tested the generalization of the proposed model with new experiments in which data from year 2015 was used for training and year 2016 for validation.  From these new set of experiments, we observed that the model, yet failing for wheat (R$^2$ $<$ 0.1), is able to resolve about 50-60\% of the variability in the rest of experiments (R$^2$  $\sim$ 0.6 for all crops, R$^2$  $\sim$ 0.5 for corn and soybean). These results can be explained by the different statistical characteristics of the two years of data we have (see Fig. \ref{fig:histograms2016}), for which the model cannot obviously extrapolate/adapt. It should be noted that studies aiming at forecasting crop yield are generally based on 5-10 years of data and adapt the feature representation and model to temporal changes, and generally reach performances of R$^2$  $\sim$ 0.7 in the best case after using crop masks \citep{kastens05, bolton13, johnson14}. In this regard, our results suggest that the proposed model trained with multi-year optical and microwave data (thus more data and more informative and complementary features) could lead to improved estimates and possibly have forecasting capabilities in forecasting settings as well. This is certainly a matter of future research.}

\section{Conclusions}\label{sec:conclusions}

The estimation of crop yield is of paramount relevance in current times due to the high pressure on food production. Satellite remotely sensed data during the growing season allows supervising the status and variation of crop conditions and parameters. Specifically, optical data captures information on photosynthetic activity (i.e., greenness, represented by EVI), and low-frequency passive microwaves provide information on plant water storage (i.e., biomass and water content, represented by VOD).
The problem is nevertheless very challenging because optical sensors are hampered by the time resolution while microwave sensor data by the spatial resolution. In this work we proposed a combination of optical and microwave data to estimate the crop yield in US Corn Belt based on (1) a new synergistic metric and (2) nonlinear machine learning regression algorithms.

We presented a new synergistic EVI/VOD lag metric as a suitable tool for crop yield assessment, and showed that it outperforms common single-sensor metrics. We demonstrated that different input measurements coming from optical and microwave sensors, EVI and VOD respectively, provide complementary information, and that the combination of both improves the estimations obtained by linear and nonlinear regression models. The use of multi-sensor data with different spectral ranges improves the crop yield estimations compared to the use of each sensor separately.
Furthermore, although seasonal metrics are practical and widely used, we show that best results are obtained if all relations and higher-order statistical moments of the full satellite time series are exploited through machine learning methods. 

We tested the performance of two machine learning regression models, a linear and a nonlinear kernel-based regression (RLR, KRR), and explored two different settings: estimation and within-season forecasting.
For both settings, we developed models for the total crop yield without differentiating crop types, and for crop-specific yields of corn, soybean, and wheat separately. The estimations obtained by the models agree well with the USDA-NASS reported survey data. The nonlinear method, KRR, provided superior accuracy in predicting yield in all experiments. The better results obtained by nonlinear models is an indicator the nonlinear relationship between the output variables and the predictors. Forecasting of total production and crop yield was possible up to four months before harvest, when models reached a plateau in accuracy. The results of the models developed for year 2015 are as good as the obtained for years 2015 and 2016 in a multi-year setting, confirming the generalization of the proposed approach. 

Our results show that homogeneous counties having a dominant crop type are not representative of the crop mixture across the study area and are more difficult to model, resulting in greater prediction errors. We suspect that aggregated time series per county are highly influenced by the presence or absence of corn, since it is the crop with highest production. The reported results suggest that using a combination of EVI and VOD has a significant benefit in prediction accuracy, even with models built to estimate the yield of minority crops such as wheat.

This study shows the capacity of VOD to complement the information from optical satellite imagery, which is currently used in most global agricultural monitoring systems \citep{Fritz2019}. The use of low-frequency passive microwave EO information alongside well-known optical indices can potentially improve present agricultural monitoring and yield prediction capabilities. %

This work tried to exploit multi-sensor fusion in different ways, and opens the door to more focused studies in the future. We foresee three lines of further research. First, other complementary sensor data could be included easily in our ML scheme such as sun-induced fluorescense signal and gross primary productivity. Second, extending the study to include episodes of vegetation water stress and disturbances appears to be a must to corroborate our results. This will also allow further insight on the physics behind the relation of the EVI/VOD lag to crop yield. Finally, other more advanced machine learning algorithms based on deep neural networks could be deployed to exploit spatio-temporal data structures.

\section*{Acknowledgments}

The research leading to these results has received funding from the European Research Council (ERC) under the ERC-CoG-2014 SEDAL under grant agreement 647423, and the Spanish Ministry of Economy and Competitiveness (MINECO) and FEDER co-funding through the projects TIN2012-38102-C03-01, TIN2015-64210-R and RTI2018-096765-A-100. MP is supported by a Ram\'on y Cajal contract (Spanish Ministry of Science, Innovation and Universities).

\bibliographystyle{elsarticle-harv}
\bibliography{erc}

\begin{thebibliography}{75}
\expandafter\ifx\csname natexlab\endcsname\relax\def\natexlab#1{#1}\fi
\expandafter\ifx\csname url\endcsname\relax
  \def\url#1{\texttt{#1}}\fi
\expandafter\ifx\csname urlprefix\endcsname\relax\def\urlprefix{URL }\fi

\bibitem[{Adsuara et~al.(2019)Adsuara, P{\'e}rez-Suay, Mu{\~n}oz-Mar\'{\i},
  Mateo-Sanchis, Piles, and Camps-Valls}]{adsuara19}
Adsuara, J.~E., P{\'e}rez-Suay, A., Mu{\~n}oz-Mar\'{\i}, J., Mateo-Sanchis, A.,
  Piles, M., Camps-Valls, G., 2019. Nonlinear distribution regression for
  remote sensing applications. IEEE Transactions on Geoscience and Remote
  Sensing.

\bibitem[{Akaike(1969)}]{Akaike69}
Akaike, H., 1969. Fitting autoregressive models for prediction. Annals of the
  Institute of Statistical Mathematics 21~(1), 243--247.

\bibitem[{Alemu and Henebry(2017)}]{alemu17}
Alemu, W.~G., Henebry, G.~M., 2017. Comparing passive microwave with
  visible-to-near-infrared phenometrics in croplands of {N}orthern {E}urasia.
  Remote Sensing 9~(6), 613.

\bibitem[{Atzberger(2013)}]{atzberger13}
Atzberger, C., 2013. Advances in remote sensing of agriculture: {Context}
  description, existing operational monitoring systems and major information
  needs. Remote Sensing 5~(2), 949--981.

\bibitem[{Bauer et~al.(1981)Bauer, Daughtry, and Vanderbilt}]{bauer81}
Bauer, M.~E., Daughtry, C., Vanderbilt, V., 1981. Spectral-agronomic
  relationships of corn, soybean and wheat canopies.

\bibitem[{Bolton and Friedl(2013)}]{bolton13}
Bolton, D.~K., Friedl, M.~A., 2013. Forecasting crop yield using remotely
  sensed vegetation indices and crop phenology metrics. Agricultural and Forest
  Meteorology 173, 74--84.

\bibitem[{Camps-Valls and Bruzzone(2009{\natexlab{a}})}]{CampsValls09wiley}
Camps-Valls, G., Bruzzone, L. (Eds.), Dec 2009{\natexlab{a}}. Kernel methods
  for Remote Sensing Data Analysis. Wiley \& Sons, UK.

\bibitem[{Camps-Valls and Bruzzone(2009{\natexlab{b}})}]{Camps-Valls2009b}
Camps-Valls, G., Bruzzone, L., 2009{\natexlab{b}}. Kernel Methods for Remote
  Sensing Data Analysis. John Wiley and Sons.

\bibitem[{Camps-Valls et~al.(2012)Camps-Valls, Mu\~noz Mar\'i, G\'omez-Chova,
  Guanter, and Calbet}]{Camps-Valls20121759}
Camps-Valls, G., Mu\~noz Mar\'i, J., G\'omez-Chova, L., Guanter, L., Calbet,
  X., 2012. Nonlinear statistical retrieval of atmospheric profiles from
  metop-iasi and mtg-irs infrared sounding data. IEEE Transactions on
  Geoscience and Remote Sensing 50~(5 PART 2), 1759--1769, cited By 9.

\bibitem[{Chaparro et~al.(2018)Chaparro, Piles, Vall-llossera, Camps, Konings,
  and Entekhabi}]{chaparro18}
Chaparro, D., Piles, M., Vall-llossera, M., Camps, A., Konings, A.~G.,
  Entekhabi, D., Jun. 2018. L-band vegetation optical depth seasonal metrics
  for crop yield assessment. Remote Sensing of Environment 212, 249--259.

\bibitem[{Chen et~al.(2018)Chen, Lu, Moran, Batistella, Dutra, Sanches,
  da~Silva, Huang, Luiz, and de~Oliveira}]{Chen18}
Chen, Y., Lu, D., Moran, E., Batistella, M., Dutra, L.~V., Sanches, I.~D.,
  da~Silva, R. F.~B., Huang, J., Luiz, A. J.~B., de~Oliveira, M. A.~F., Jul.
  2018. Mapping croplands, cropping patterns, and crop types using {MODIS}
  time-series data. International Journal of Applied Earth Observation and
  Geoinformation 69, 133--147.

\bibitem[{Clevers and Van~Leeuwen(1996)}]{clevers96}
Clevers, J., Van~Leeuwen, H. J.~C., 1996. Combined use of optical and microwave
  remote sensing data for crop growth monitoring. Remote Sensing of Environment
  56~(1), 42--51.

\bibitem[{Doraiswamy et~al.(2003)Doraiswamy, Moulin, Cook, and
  Stern}]{doraiswamy03}
Doraiswamy, P.~C., Moulin, S., Cook, P.~W., Stern, A., 2003. Crop yield
  assessment from remote sensing. Photogrammetric engineering \& remote sensing
  69~(6), 665--674.

\bibitem[{Feldman et~al.(2018)Feldman, Short~Gianotti, Konings, McColl, Akbar,
  Salvucci, and Entekhabi}]{Feldman2018}
Feldman, A.~F., Short~Gianotti, D.~J., Konings, A.~G., McColl, K.~A., Akbar,
  R., Salvucci, G.~D., Entekhabi, D., 2018. Moisture pulse-reserve in the
  soil-plant continuum observed across biomes. Nature Plants 4~(12),
  1026--1033.

\bibitem[{Fieuzal et~al.(2017)Fieuzal, Sicre, and Baup}]{fieuzal17}
Fieuzal, R., Sicre, C.~M., Baup, F., 2017. Estimation of corn yield using
  multi-temporal optical and radar satellite data and artificial neural
  networks. International journal of applied earth observation and
  geoinformation 57, 14--23.

\bibitem[{Foley et~al.(2011)Foley, Ramankutty, Brauman, Cassidy, Gerber,
  Johnston, Mueller, O’Connell, Ray, and West}]{foley11}
Foley, J.~A., Ramankutty, N., Brauman, K.~A., Cassidy, E.~S., Gerber, J.~S.,
  Johnston, M., Mueller, N.~D., O’Connell, C., Ray, D.~K., West, P.~C., 2011.
  Solutions for a cultivated planet. Nature 478~(7369), 337.

\bibitem[{Fritz et~al.(2019)Fritz, See, Bayas, Waldner, Jacques, Becker-Reshef,
  Whitcraft, Baruth, Bonifacio, Crutchfield, Rembold, Rojas, Schucknecht, der
  Velde, Verdin, Wu, Yan, You, Gilliams, Mücher, Tetrault, Moorthy, and
  McCallum}]{Fritz2019}
Fritz, S., See, L., Bayas, J. C.~L., Waldner, F., Jacques, D., Becker-Reshef,
  I., Whitcraft, A., Baruth, B., Bonifacio, R., Crutchfield, J., Rembold, F.,
  Rojas, O., Schucknecht, A., der Velde, M.~V., Verdin, J., Wu, B., Yan, N.,
  You, L., Gilliams, S., Mücher, S., Tetrault, R., Moorthy, I., McCallum, I.,
  2019. A comparison of global agricultural monitoring systems and current
  gaps. Agricultural Systems 168, 258 -- 272.

\bibitem[{Gonz\'{a}lez-Sanchez et~al.(2014)Gonz\'{a}lez-Sanchez, Frausto-Solis,
  and Ojeda}]{gonzalez14}
Gonz\'{a}lez-Sanchez, A., Frausto-Solis, J., Ojeda, W., Jun. 2014. Predictive
  ability of machine learning methods for massive crop yield prediction.
  SPANISH JOURNAL OF AGRICULTURAL RESEARCH.

\bibitem[{Grant et~al.(2016)Grant, Wigneron, De~Jeu, Lawrence, Mialon,
  Richaume, Al~Bitar, Drusch, Van~Marle, and Kerr}]{grant16}
Grant, J.~P., Wigneron, J.-P., De~Jeu, R. A.~M., Lawrence, H., Mialon, A.,
  Richaume, P., Al~Bitar, A., Drusch, M., Van~Marle, M. J.~E., Kerr, Y., 2016.
  Comparison of {SMOS} and {AMSR}-{E} vegetation optical depth to four
  {MODIS}-based vegetation indices. Remote Sensing of Environment 172, 87--100.

\bibitem[{Guan et~al.(2017)Guan, Wu, Kimball, Anderson, Frolking, Li, Hain, and
  Lobell}]{guan17}
Guan, K., Wu, J., Kimball, J.~S., Anderson, M.~C., Frolking, S., Li, B., Hain,
  C.~R., Lobell, D.~B., Sep. 2017. The shared and unique values of optical,
  fluorescence, thermal and microwave satellite data for estimating large-scale
  crop yields. Remote Sensing of Environment 199, 333--349.

\bibitem[{Hill and Donald(2003)}]{hill2003}
Hill, M.~J., Donald, G.~E., 2003. Estimating spatio-temporal patterns of
  agricultural productivity in fragmented landscapes using avhrr ndvi time
  series. Remote Sensing of Environment 84~(3), 367--384.

\bibitem[{Hofmann et~al.(2008)Hofmann, Sch{\"o}lkopf, and Smola}]{hofmann2008}
Hofmann, T., Sch{\"o}lkopf, B., Smola, A.~J., 2008. Kernel methods in machine
  learning. The annals of statistics, 1171--1220.

\bibitem[{Hornbuckle et~al.(2016)Hornbuckle, Patton, VanLoocke, Suyker, Roby,
  Walker, Iyer, Herzmann, and Endacott}]{Hornbuckle2016}
Hornbuckle, B.~K., Patton, J.~C., VanLoocke, A., Suyker, A.~E., Roby, M.~C.,
  Walker, V.~A., Iyer, E.~R., Herzmann, D.~E., Endacott, E.~A., 2016. {SMOS}
  optical thickness changes in response to the growth and development of crops,
  crop management, and weather. Remote Sensing of Environment 180, 320 -- 333.

\bibitem[{Huete et~al.(2002)Huete, Didan, Miura, Rodriguez, Gao, and
  Ferreira}]{Huete2002}
Huete, A., Didan, K., Miura, T., Rodriguez, E., Gao, X., Ferreira, L., 2002.
  Overview of the radiometric and biophysical performance of the modis
  vegetation indices. Remote Sensing of Environment 83~(1), 195 -- 213, the
  Moderate Resolution Imaging Spectroradiometer (MODIS): a new generation of
  Land Surface Monitoring.

\bibitem[{Idso et~al.(1977)Idso, Jackson, and Reginato}]{idso77}
Idso, S.~B., Jackson, R.~D., Reginato, R.~J., 1977. Remote-sensing of crop
  yields. Science 196~(4285), 19--25.

\bibitem[{Jackson and Schmugge(1991)}]{Jackson1991}
Jackson, T., Schmugge, T., 1991. Vegetation effects on the microwave emission
  of soils. Remote Sensing of Environment 36~(3), 203 -- 212.

\bibitem[{Johnson(2014)}]{johnson14}
Johnson, D.~M., 2014. An assessment of pre-and within-season remotely sensed
  variables for forecasting corn and soybean yields in the {United} {States}.
  Remote Sensing of Environment 141, 116--128.

\bibitem[{Jones et~al.(2011)Jones, Jones, Kimball, and McDonald}]{jones11}
Jones, M.~O., Jones, L.~A., Kimball, J.~S., McDonald, K.~C., 2011. Satellite
  passive microwave remote sensing for monitoring global land surface
  phenology. Remote Sensing of Environment 115~(4), 1102--1114.

\bibitem[{Kastens et~al.(2005)Kastens, Kastens, Kastens, Price, Martinko, and
  Lee}]{kastens05}
Kastens, J.~H., Kastens, T.~L., Kastens, D.~L., Price, K.~P., Martinko, E.~A.,
  Lee, R.-Y., 2005. Image masking for crop yield forecasting using avhrr ndvi
  time series imagery. Remote Sensing of Environment 99~(3), 341--356.

\bibitem[{Kern et~al.(2018)Kern, Barcza, Marjanović, Árendás, Fodor, Bónis,
  Bognár, and Lichtenberger}]{Kern18}
Kern, A., Barcza, Z., Marjanović, H., Árendás, T., Fodor, N., Bónis, P.,
  Bognár, P., Lichtenberger, J., Oct. 2018. Statistical modelling of crop
  yield in {Central} {Europe} using climate data and remote sensing vegetation
  indices. Agricultural and Forest Meteorology 260-261, 300--320.

\bibitem[{Konings et~al.(2017)Konings, Piles, Das, and Entekhabi}]{Konings2017}
Konings, A.~G., Piles, M., Das, N., Entekhabi, D., 2017. L-band vegetation
  optical depth and effective scattering albedo estimation from {SMAP}. Remote
  Sensing of Environment 198, 460 -- 470.

\bibitem[{Konings et~al.(2016)Konings, Piles, Rötzer, McColl, Chan, and
  Entekhabi}]{Konings2016}
Konings, A.~G., Piles, M., Rötzer, K., McColl, K.~A., Chan, S.~K., Entekhabi,
  D., 2016. Vegetation optical depth and scattering albedo retrieval using time
  series of dual-polarized l-band radiometer observations. Remote Sensing of
  Environment 172, 178 -- 189.

\bibitem[{Lawrence et~al.(2014)Lawrence, Wigneron, Richaume, Novello, Grant,
  Mialon, Al~Bitar, Merlin, Guyon, and Leroux}]{lawrence14}
Lawrence, H., Wigneron, J.-P., Richaume, P., Novello, N., Grant, J., Mialon,
  A., Al~Bitar, A., Merlin, O., Guyon, D., Leroux, D., 2014. Comparison between
  {SMOS} {Vegetation} {Optical} {Depth} products and {MODIS} vegetation indices
  over crop zones of the {USA}. Remote Sensing of Environment 140, 396--406.

\bibitem[{Lewis-Beck et~al.(2018)Lewis-Beck, Niemi, Caragea, Hornbuckle, and
  Walker}]{lewis_beck_18}
Lewis-Beck, C., Niemi, J., Caragea, P., Hornbuckle, B., Walker, V., 2018.
  Monitoring {Crop} {Growth} in the {Us} {Corn} {Belt} with {SMOS} {Level} 2
  {Tau}. In: 2018 {IEEE} 15th {Specialist} {Meeting} on {Microwave}
  {Radiometry} and {Remote} {Sensing} of the {Environment} ({MicroRad}). IEEE,
  pp. 1--4.

\bibitem[{Li et~al.(2007)Li, Liang, Wang, and Qin}]{li07}
Li, A., Liang, S., Wang, A., Qin, J., 2007. Estimating crop yield from
  multi-temporal satellite data using multivariate regression and neural
  network techniques. Photogrammetric Engineering \& Remote Sensing 73~(10),
  1149--1157.

\bibitem[{Li et~al.(2017)Li, Low, Lobell, and Ermon}]{Xiaocheng17}
Li, X., Low, M., Lobell, D.~B., Ermon, S., 2017. Deep {Gaussian} {Process} for
  {Crop} {Yield} {Prediction} {Based} on {Remote} {Sensing} {Data}.

\bibitem[{Liu et~al.(2011)Liu, de~Jeu, McCabe, Evans, and van Dijk}]{Liu_2011}
Liu, Y.~Y., de~Jeu, R.~A., McCabe, M.~F., Evans, J.~P., van Dijk, A.~I., 2011.
  Global long-term passive microwave satellite-based retrievals of vegetation
  optical depth. Geophysical Research Letters 38~(18).

\bibitem[{Lobell and Field(2007)}]{lobell07}
Lobell, D.~B., Field, C.~B., 2007. Global scale climate–crop yield
  relationships and the impacts of recent warming. Environmental Research
  Letters 2~(1), 014002.

\bibitem[{Long and Ulaby(2014)}]{Ulaby2014}
Long, D., Ulaby, F.~T., 2014. {Microwave remote sensing active and passive}.
  The University of Michigan Press.

\bibitem[{L{\'o}pez-Lozano et~al.(2015{\natexlab{a}})L{\'o}pez-Lozano,
  Duveiller, Seguini, Meroni, Garc{\'\i}a-Condado, Hooker, Leo, and
  Baruth}]{lopez15}
L{\'o}pez-Lozano, R., Duveiller, G., Seguini, L., Meroni, M.,
  Garc{\'\i}a-Condado, S., Hooker, J., Leo, O., Baruth, B., 2015{\natexlab{a}}.
  Towards regional grain yield forecasting with 1 km-resolution eo biophysical
  products: Strengths and limitations at pan-european level. Agricultural and
  forest meteorology 206, 12--32.

\bibitem[{L{\'o}pez-Lozano et~al.(2015{\natexlab{b}})L{\'o}pez-Lozano,
  Duveiller, Seguini, Meroni, García-Condado, Hooker, Leo, and
  Baruth}]{LopezLozano15}
L{\'o}pez-Lozano, R., Duveiller, G., Seguini, L., Meroni, M., García-Condado,
  S., Hooker, J., Leo, O., Baruth, B., 2015{\natexlab{b}}. Towards regional
  grain yield forecasting with 1km-resolution eo biophysical products:
  Strengths and limitations at pan-european level. Agricultural and Forest
  Meteorology 206, 12 -- 32.

\bibitem[{MacDonald and Hall(1980)}]{macdonald80}
MacDonald, R.~B., Hall, F.~G., 1980. Global crop forecasting. Science
  208~(4445), 670--679.

\bibitem[{Marinković et~al.(2009)Marinković, Crnobarac, Brdar, Antić,
  Jaćimović, and Crnojević}]{marinkovic09}
Marinković, B., Crnobarac, J., Brdar, S., Antić, B., Jaćimović, G.,
  Crnojević, V., 2009. Data mining approach for predictive modeling of
  agricultural yield data. In: Proc. {First} {Int} {Workshop} on {Sensing}
  {Technologies} in {Agriculture}, {Forestry} and {Environment} ({BioSense}09),
  {Novi} {Sad}, {Serbia}. pp. 1--5.

\bibitem[{Mateo-Sagasta et~al.(2018)Mateo-Sagasta, Marjani~Zadeh, and
  Turral}]{mateo_sagasta18}
Mateo-Sagasta, J., Marjani~Zadeh, S., Turral, H., 2018. More people, more
  food… worse water? - {Water} {Pollution} from {Agriculture}: a global
  review. FAO, Rome, Italy.

\bibitem[{Mkhabela et~al.(2011)Mkhabela, Bullock, Raj, Wang, and
  Yang}]{mkhabela11}
Mkhabela, M., Bullock, P., Raj, S., Wang, S., Yang, Y., 2011. Crop yield
  forecasting on the canadian prairies using modis ndvi data. Agricultural and
  Forest Meteorology 151~(3), 385--393.

\bibitem[{Mladenova et~al.(2017)Mladenova, Bolten, Crow, Anderson, Hain,
  Johnson, and Mueller}]{mladenova17}
Mladenova, I.~E., Bolten, J.~D., Crow, W.~T., Anderson, M.~C., Hain, C.~R.,
  Johnson, D.~M., Mueller, R., 2017. Intercomparison of soil moisture,
  evaporative stress, and vegetation indices for estimating corn and soybean
  yields over the us. IEEE Journal of Selected Topics in Applied Earth
  Observations and Remote Sensing 10~(4), 1328--1343.

\bibitem[{Momen et~al.(2017)Momen, Wood, Novick, Pangle, Pockman, McDowell, and
  Konings}]{Momen2017}
Momen, M., Wood, J.~D., Novick, K.~A., Pangle, R., Pockman, W.~T., McDowell,
  N.~G., Konings, A.~G., 2017. Interacting effects of leaf water potential and
  biomass on vegetation optical depth. Journal of Geophysical Research:
  Biogeosciences 122~(11), 3031--3046.

\bibitem[{O'Neill et~al.(2018)O'Neill, Chan, Njoku, Jackson, and
  Bindlish}]{ONeill2018}
O'Neill, P., Chan, S., Njoku, E., Jackson, T., Bindlish, R., 2018. {SMAP}
  enhanced l2 radiometer half-orbit 9 km ease-grid soil moisture, version 1.
  Tech. Rep. https://doi.org/10.5067/K4A1SNL5DLON, NASA National Snow and Ice
  Data Center Distributed Active Archive Center. Boulder, Colorado USA.

\bibitem[{Patton and Hornbuckle(2013)}]{patton13}
Patton, J., Hornbuckle, B., 2013. Initial validation of {SMOS} vegetation
  optical thickness in {Iowa}. IEEE Geoscience and Remote Sensing Letters
  10~(4), 647--651.

\bibitem[{Piles et~al.(2017)Piles, Camps-Valls, Chaparro, Entekhabi, Konings,
  and Jagdhuber}]{piles17}
Piles, M., Camps-Valls, G., Chaparro, D., Entekhabi, D., Konings, A.~G.,
  Jagdhuber, T., 2017. Remote sensing of vegetation dynamics in agro-ecosystems
  using {SMAP} vegetation optical depth and optical vegetation indices. In:
  2017 IEEE International Geoscience and Remote Sensing Symposium (IGARSS).
  IEEE, pp. 4346--4349.

\bibitem[{Prasad et~al.(2006)Prasad, Chai, Singh, and Kafatos}]{Prasad06}
Prasad, A.~K., Chai, L., Singh, R.~P., Kafatos, M., 2006. Crop yield estimation
  model for {Iowa} using remote sensing and surface parameters. International
  Journal of Applied Earth Observation and Geoinformation 8~(1), 26--33.

\bibitem[{Quarmby et~al.(1993)Quarmby, Milnes, Hindle, and Silleos}]{quarmby93}
Quarmby, N.~A., Milnes, M., Hindle, T.~L., Silleos, N., 1993. The use of
  multi-temporal {NDVI} measurements from {AVHRR} data for crop yield
  estimation and prediction. International Journal of Remote Sensing 14~(2),
  199--210.

\bibitem[{Reed et~al.(1994)Reed, Brown, VanderZee, Loveland, Merchant, and
  Ohlen}]{reed94}
Reed, B.~C., Brown, J.~F., VanderZee, D., Loveland, T.~R., Merchant, J.~W.,
  Ohlen, D.~O., 1994. Measuring phenological variability from satellite
  imagery. Journal of vegetation science 5~(5), 703--714.

\bibitem[{Ren et~al.(2008)Ren, Chen, Zhou, and Tang}]{ren08}
Ren, J., Chen, Z., Zhou, Q., Tang, H., 2008. Regional yield estimation for
  winter wheat with {MODIS}-{NDVI} data in {Shandong}, {China}. International
  Journal of Applied Earth Observation and Geoinformation 10~(4), 403--413.

\bibitem[{Rojo-\'Alvarez et~al.(2018)Rojo-\'Alvarez, Mart\'inez-Ram\'on,
  Mu{\~n}oz-Mar\'i, and Camps-Valls}]{Rojo17dspkm}
Rojo-\'Alvarez, J., Mart\'inez-Ram\'on, M., Mu{\~n}oz-Mar\'i, J., Camps-Valls,
  G., Ap 2018. Digital Signal Processing with Kernel Methods. Wiley \& Sons,
  UK.

\bibitem[{Ruescas et~al.(2018)Ruescas, Hieronymi, Mateo-Garcia, Koponen,
  Kallio, and Camps-Valls}]{rs10050786}
Ruescas, A.~B., Hieronymi, M., Mateo-Garcia, G., Koponen, S., Kallio, K.,
  Camps-Valls, G., 2018. Machine learning regression approaches for colored
  dissolved organic matter (cdom) retrieval with s2-msi and s3-olci simulated
  data. Remote Sensing 10~(5).

\bibitem[{Sakamoto et~al.(2013)Sakamoto, Gitelson, and Arkebauer}]{sakamoto13}
Sakamoto, T., Gitelson, A.~A., Arkebauer, T.~J., 2013. {MODIS}-based corn grain
  yield estimation model incorporating crop phenology information. Remote
  Sensing of Environment 131, 215--231.

\bibitem[{Sakamoto et~al.(2014)Sakamoto, Gitelson, and Arkebauer}]{sakamoto14}
Sakamoto, T., Gitelson, A.~A., Arkebauer, T.~J., 2014. Near real-time
  prediction of {US} corn yields based on time-series {MODIS} data. Remote
  Sensing of Environment 147, 219--231.

\bibitem[{Schnitkey(2013)}]{schnitkey13}
Schnitkey, G., Jul. 2013. Concentration of {Corn} and {Soybean} {Production} in
  the {U}.{S}. • farmdoc daily.

\bibitem[{Shawe-Taylor and Cristianini(2004)}]{Shawetaylor04}
Shawe-Taylor, J., Cristianini, N., 2004. Kernel {M}ethods for {P}attern
  {A}nalysis. Cambridge University Press.

\bibitem[{Son et~al.(2014)Son, Chen, Chen, Minh, and Trung}]{son14}
Son, N.~T., Chen, C.~F., Chen, C.~R., Minh, V.~Q., Trung, N.~H., Oct. 2014. A
  comparative analysis of multitemporal {MODIS} {EVI} and {NDVI} data for
  large-scale rice yield estimation. Agricultural and Forest Meteorology 197,
  52--64.

\bibitem[{Theodoridis and Koutroumbas(2009)}]{theodoridis2009}
Theodoridis, S., Koutroumbas, K., 2009. Pattern recognition. 2003. Elsevier
  Inc.

\bibitem[{Tian et~al.(2016)Tian, Brandt, Liu, Verger, Tagesson, Diouf,
  Rasmussen, Mbow, Wang, and Fensholt}]{tian16}
Tian, F., Brandt, M., Liu, Y.~Y., Verger, A., Tagesson, T., Diouf, A.~A.,
  Rasmussen, K., Mbow, C., Wang, Y., Fensholt, R., 2016. Remote sensing of
  vegetation dynamics in drylands: Evaluating vegetation optical depth (vod)
  using avhrr ndvi and in situ green biomass data over west african sahel.
  Remote sensing of environment 177, 265--276.

\bibitem[{Tian et~al.(2018)Tian, Wigneron, Ciais, Chave, Og{\'e}e,
  Pe{\~n}uelas, R{\ae}bild, Domec, Tong, Brandt, et~al.}]{tian18}
Tian, F., Wigneron, J.-P., Ciais, P., Chave, J., Og{\'e}e, J., Pe{\~n}uelas,
  J., R{\ae}bild, A., Domec, J.-C., Tong, X., Brandt, M., et~al., 2018.
  Coupling of ecosystem-scale plant water storage and leaf phenology observed
  by satellite. Nature ecology \& evolution 2~(9), 1428.

\bibitem[{Tramontana et~al.(2016)Tramontana, Jung, Camps-Valls, Ichii, Raduly,
  Reichstein, Schwalm, Arain, Cescatti, Kiely, Merbold, Serrano-Ortiz, Sickert,
  Wolf, and Papale}]{Tramontana16bg}
Tramontana, G., Jung, M., Camps-Valls, G., Ichii, K., Raduly, B., Reichstein,
  M., Schwalm, C.~R., Arain, M.~A., Cescatti, A., Kiely, G., Merbold, L.,
  Serrano-Ortiz, P., Sickert, S., Wolf, S., Papale, D., 2016. Predicting carbon
  dioxide and energy fluxes across global fluxnet sites with regression
  algorithms. Biogeosciences Discussions 2016, 1--33.

\bibitem[{USDA(1992)}]{usda1992weights}
USDA, 1992. Weights, measures, and conversion factors for agricultural
  commodities and their products.

\bibitem[{Van~Wart et~al.(2013)Van~Wart, Kersebaum, Peng, Milner, and
  Cassman}]{wart13}
Van~Wart, J., Kersebaum, K.~C., Peng, S., Milner, M., Cassman, K.~G., 2013.
  Estimating crop yield potential at regional to national scales. Field Crops
  Research 143, 34--43.

\bibitem[{Wigneron et~al.(2017)Wigneron, Jackson, O'Neill, Lannoy, de~Rosnay,
  Walker, Ferrazzoli, Mironov, Bircher, Grant, Kurum, Schwank, noz Sabater,
  Das, Royer, Al-Yaari, Bitar, Fernandez-Moran, Lawrence, Mialon, Parrens,
  Richaume, Delwart, and Kerr}]{Wigneron2017}
Wigneron, J.-P., Jackson, T., O'Neill, P., Lannoy, G.~D., de~Rosnay, P.,
  Walker, J., Ferrazzoli, P., Mironov, V., Bircher, S., Grant, J., Kurum, M.,
  Schwank, M., noz Sabater, J.~M., Das, N., Royer, A., Al-Yaari, A., Bitar,
  A.~A., Fernandez-Moran, R., Lawrence, H., Mialon, A., Parrens, M., Richaume,
  P., Delwart, S., Kerr, Y., 2017. Modelling the passive microwave signature
  from land surfaces: A review of recent results and application to the l-band
  {SMOS} \& {SMAP} soil moisture retrieval algorithms. Remote Sensing of
  Environment 192, 238 -- 262.

\bibitem[{Xin et~al.(2013)Xin, Gong, Yu, Yu, Broich, Suyker, and
  Myneni}]{xin13}
Xin, Q., Gong, P., Yu, C., Yu, L., Broich, M., Suyker, A.~E., Myneni, R.~B.,
  2013. A production efficiency model-based method for satellite estimates of
  corn and soybean yields in the midwestern us. Remote Sensing 5~(11),
  5926--5943.

\bibitem[{You et~al.(2017)You, Li, Low, Lobell, and Ermon}]{you17}
You, J., Li, X., Low, M., Lobell, D., Ermon, S., 2017. Deep gaussian process
  for crop yield prediction based on remote sensing data. In: AAAI. pp.
  4559--4566.

\bibitem[{Zhang et~al.(2003{\natexlab{a}})Zhang, Wu, and Liu}]{zhang03}
Zhang, F., Wu, B., Liu, C., 2003{\natexlab{a}}. Using time series of spot vgt
  ndvi for crop yield forecasting. In: Geoscience and Remote Sensing Symposium,
  2003. IGARSS'03. Proceedings. 2003 IEEE International. Vol.~1. Ieee, pp.
  386--388.

\bibitem[{Zhang et~al.(2014)Zhang, Feng, and Yao}]{zhang14}
Zhang, J., Feng, L., Yao, F., Aug. 2014. Improved maize cultivated area
  estimation over a large scale combining {MODIS} {EVI} time series data and
  crop phenological information. ISPRS Journal of Photogrammetry and Remote
  Sensing 94, 102--113.

\bibitem[{Zhang et~al.(2010)Zhang, Anderson, Tan, Barlow, and Myneni}]{zhang10}
Zhang, P., Anderson, B., Tan, B., Barlow, M., Myneni, R., 2010. Monitoring crop
  yield in usa using a satellite-based climate-variability impact index. In:
  Geoscience and Remote Sensing Symposium (IGARSS), 2010 IEEE International.
  IEEE, pp. 1815--1818.

\bibitem[{Zhang et~al.(2003{\natexlab{b}})Zhang, Friedl, Schaaf, Strahler,
  Hodges, Gao, Reed, and Huete}]{zhang2003}
Zhang, X., Friedl, M.~A., Schaaf, C.~B., Strahler, A.~H., Hodges, J.~C., Gao,
  F., Reed, B.~C., Huete, A., 2003{\natexlab{b}}. Monitoring vegetation
  phenology using {MODIS}. Remote sensing of environment 84~(3), 471--475.

\bibitem[{Zhong et~al.(2016)Zhong, Yu, Li, Hu, and Gong}]{zhong16}
Zhong, L., Yu, L., Li, X., Hu, L., Gong, P., Nov. 2016. Rapid corn and soybean
  mapping in {US} {Corn} {Belt} and neighboring areas. Scientific Reports 6,
  36240.

\end{thebibliography}

\section*{List of Figure Captions}
Figure 1: Study Area including the 8 states and cropland mask following the MODIS IGBP land cover classification. Only pixels classified as croplands were considered in this study.

Figure 2: Left: histogram on the reported yield per county for the three major crops grown in the study area (soybean, wheat and corn). Right: County-scale percentages of area planted for corn, soybean, and other crops (mainly wheat). Each vertex in the triangle corresponds to 100\%

Figure 3: Schematic representation of the proposed methodology: a) pre-processing steps, b) set-up of input feature space, and c) crop yield assessment. We start by extracting the MODIS EVI and SMAP VOD time series at the county scale. These time series are then used to capture crop phenology and predict yield in two complementary ways: using summarizing metrics (i.e., maximum value, small integral, PC1 or EVI/VOD lag) or the full temporal series. Subsequently, these descriptors are fed into a machine learning regression algorithm and trained to estimate crop yield. After training the models with a reduced set, they are applied to the whole area to generate spatially explicit crop yield estimates.

Figure 4: Left: County-scaled time series of MODIS EVI and SMAP VOD for the study area (see Fig.1). Mean values are overlaid. Right: scatter plot showing the relationship of EVI/VOD lag and the associated yield per county. Colours represent the percentage of area planted for corn, soybean, and other crops (mainly wheat). Each vertex in the triangle corresponds to 100\%.

Figure 5: County-level maps of yield survey data [$kg/m^2$] (left), best yield estimation obtained with full time series and KRR [$kg/m^2$] (in bold text on Tables \ref{tab:resultsall} and \ref{tab:resultscrops}) (middle) and relative error [\%] (right). From top to bottom: total yield, corn, soybean and wheat crop yields.

Figure 6: Scatter plots showing the residuals in the estimation of crop yield ($y-\hat{y}$) [kg$\cdot$m$^{-2}$] vs. the survey yield for the best prediction model and the four sceanrios: total yield, corn, soybean and wheat yield (in bold, from Tables \ref{tab:resultsall} and \ref{tab:resultscrops}). Colors depict the percentage of corn, soybean and wheat planted within each county. Dashed lines indicate the 25th-75th percentiles of the residuals distribution.

Figure 7: R$^2$ and RMSE [kg$\cdot$ m$^{-2}$] evolution as a function of the temporal window used to develop the models.

Figure 8: Top: mean time series of EVI and VOD data of both 2015 and 2016 years. Histograms plot showing the input values for specific time. Bottom: Histograms on the reported yield per county for crops scenarios with 2015 data and 2016 data.

\clearpage

\end{document}